\documentclass[letterpaper,twocolumn,10pt]{article}
\usepackage{usenix-2020-09}

\usepackage{tikz}
\usepackage{amsmath}
\usepackage{pifont}
\usepackage{subcaption}
\usepackage{multirow}
\usepackage{cite}
\usepackage{filecontents}
\usepackage{siunitx}

\begin{document}
%-------------------------------------------------------------------------------

%don't want date printed
\date{}

% make title bold and 14 pt font (Latex default is non-bold, 16 pt)
\title{\Large \bf TimeTravel: Real-time Timing Drift Attack on System Time Using Acoustic Waves}

%for single author (just remove % characters)
% \author{
% {\rm Jianshuo Liu}\\
% Institute of Information Engineering, Chinese Academy of Sciences
% \and
% {\rm Hong Li}\\
% Second Institution
% % copy the following lines to add more authors
% % \and
% % {\rm Name}\\
% %Name Institution
% } % end author

\author{Jianshuo Liu$^{1,2}$, Hong Li$^{1,2}$\thanks{Hong Li is corresponding author.}, Haining Wang$^{3}$, Mengjie Sun$^{1,2}$, Hui Wen$^{1,2}$, Jinfa Wang$^{1,2}$, Limin Sun$^{1,2}$
    \\ $^{1}$ Institute of Information Engineering, Chinese Academy of Sciences
    \\ $^{2}$ School of Cyber Security, University of Chinese Academy of Sciences
    \\ $^{3}$ Department of Electrical and Computer Engineering, Virginia Tech
    \\\{liujianshuo,lihong,sunmengjie,wenhui,wangjinfa,sunlimin\}@iie.ac.cn,hnw@vt.edu
    % \thanks{$^{\dag}$Hong Li is corresponding author.}
    % \thanks{E-mail: \{liujianshuo,lihong,sunmengjie,wenhui,lizhi,sunlimin\}@iie.ac.cn,}
    % \thanks{hnw@vt.edu}
}

\maketitle

%-------------------------------------------------------------------------------
\begin{abstract}
%-------------------------------------------------------------------------------
Real-time Clock (RTC) has been widely used in various real-time systems to provide precise system time. In this paper, we reveal a new security vulnerability of the RTC circuit, where the internal storage time or timestamp can be arbitrarily modified forward or backward. The security threat of dynamic modifications of system time caused by this vulnerability is called \textit{TimeTravel}. 
%By utilizing 
Based on acoustic resonance and piezoelectric effects, TimeTravel applies acoustic guide waves to the quartz crystal, thereby adjusting the characteristics of the oscillating signal transmitted into the RTC circuit. By manipulating the parameters of acoustic waves, TimeTravel can accelerate or decelerate the timing speed of system time at an adjustable rate, resulting in the relative drift of the timing, which can pose serious safety threats. To assess the severity of TimeTravel, we examine nine modules and seven commercial devices under the RTC circuit. The experimental results show that TimeTravel can drift system time forward and backward at a chosen speed with a maximum 93\% accuracy. Our analysis further shows that TimeTravel can maintain an attack success rate of no less than 77\% under environments with typical obstacle items.

\end{abstract}

%-------------------------------------------------------------------------------
\section{Introduction}
%-------------------------------------------------------------------------------

% Maintaining clock stability in computing and interconnected systems is important for the proper operation of real-time applications. In these systems, Real-Time Clock (RTC) modules play a major role. RTC time-keep

Maintaining the stability of the internal clock is crucial in real-time digital systems. In these systems, the system 
time-based signal plays an important role in keeping the clock accurate. The real-time clock (RTC) is a kind of simple, low-power circuit consisting of several counters widely used in electronic devices to provide time base signals\cite{goldband1979real}. Due to its low power consumption, easy to be integrated, and precise timing characteristics, RTC has become a necessity for upholding the rhythm of timing in many low-speed applications, such as scheduling tasks in industrial automation, health measuring by healthcare devices, providing interruption for smart vehicle devices, and ensuring transaction accuracy in financial systems. 

Typically, RTC is usually soldered onto a PCB board as an independent module or is embedded inside a microcontroller as an integrated timing unit\cite{khan2012designing}.
For clock timing applications, RTC requires users to manually set the initial time (or use the NTP protocol\cite{rytilahti2018masters} to obtain the time) and then store it in the counter. Afterward, RTC generates 1-second time base signals (calendar mode) and communicates with other components through the system bus. For timestamp timing applications, RTC generates millisecond-level timestamp signals (32-bit mode) through the collaboration of frequency dividers and counters, providing more accurate timing functionality.
When the microcontroller receives the time or timestamp, it will forward the time to higher-level applications or smart devices for further processing. 
Although obtaining system time is simple, the process may lack the necessary verification mechanism, making the time vulnerable to manipulation. Previous research has demonstrated that time synchronization request packages sent by clients can be hijacked, which in turn tampers with the client's local time\cite{ntpattack,perry2021devil}. Attackers can also slow down or speed up a Bitcoin node's network timestamp counter by connecting as multiple peers and reporting inaccurate timestamps, which may increase the chances of a successful double-spending. 

While there has been extensive exploration into %attacks of 
malicious modification of system time or timestamp, these explorations often remain at the network layer and only work under certain assumptions. One previous work\cite{ntpattack} assumes that NTP servers fragment packets to a 68-byte MTU; however, many OSes (e.g., Windows and Linux) already avoid this trait. 
Internet of Things (IoT) devices are commonly not configured with NTP, due to the limitations of timer and network performance.
Bitcoin transactions nowadays rely on authoritative timestamp servers, which can generate timestamps with digital signatures to prevent malicious tampering\cite{ma2020achieving}.  
% Moreover, network-based attacks are ineffective against devices that rely on internationally authoritative NTP servers\cite{authntpserver} for synchronization, or more generally IoT devices that even do not require an Internet connection. 
%Generally, 
%\textcolor{red}{Currently, many devices are not connected to the network, and they only rely on local timestamps or relative variances in timestamps to maintain work logic. 
In addition, some devices are not connected to the Internet, and they only rely on local timestamps or relative variances in timestamps to maintain their work logic. 
If the system time is maliciously changed, it may have serious consequences, such as property damage to the organization due to incorrect execution of equipment operations or even timing failure of medical equipment that threatens human lives. However, timing attacks under these scenarios without NTP support have not yet been well studied.

On one hand, attackers usually have limited privileges to access a target device, and thus
it is challenging to modify the time or timestamp saved by RTC counters.
%it is difficult to launch a time-hijacking attack. 
For devices that use millisecond timestamps for maintaining work logic, it is almost impossible for attackers to dynamically tamper with such fine-grained timestamps because of the inaccessibility of related interfaces. 
%However,  despite existing limitations, 
On the other hand, many devices are often exposed in public areas that attackers can access, which inspires us to explore from a different angle. Can attackers modify the system time or timestamp solely through physical channels without requiring additional privileges, while maintaining a certain physical distance from the target device?

In this paper, we explore the vulnerability of RTC concerning time modification from a new perspective. Specifically, we reveal that attackers can modify the time or timestamp inside the RTC without privileges to compromise the device or the network where the device is connected, thereby triggering a series of severe consequences. The attack vector is to apply acoustic vibrations to the medium surface (e.g., a desk) where the target device is placed. This kind of acoustic wave may cause additional resonance in the quartz crystal oscillator in the RTC circuit, and then stimulate additional electrical responses.
% The oscillator can generate stable periodic sinusoidal signals. In the RTC circuit, such signals can be used as the time base after frequency division. 
Any interference to the signal output by the oscillator may affect the accuracy of the time-base signal, thereby affecting the timing accuracy. We discover that by applying specific carefully crafted acoustic signals and propagating them to the oscillator, an attacker could instantly adjust the device's time forward or backward at different speeds. We call such a security threat \textit{TimeTravel}. 
To the best of our knowledge, this is the first work to explore the feasibility of modifying the internal system time or timestamp in RTC directly via physical interference, which may disrupt the operational logic of various devices. 
We assess the security risks of TimeTravel on nine off-the-shelf modules with the configuration of an RTC circuit, as well as seven commercial devices for real-life attack scenarios, and the results show that TimeTravel can successfully adjust the system time at different drifting rates with a maximum 93\% success rate, and achieves desired robustness under environments with physical obstacles.

The main contributions of this work are summarized below:
\begin{enumerate}
    \item We reveal a new security threat TimeTravel, which
    is the first real-time system time modification attack via a physical interference channel. TimeTravel is capable of arbitrarily drifting the time or timestamp in RTC at different speeds. 
    %To the best of our knowledge, this is the first work to explore the feasibility of modifying the internal system time or timestamp in RTC directly via physical interference, which may disrupt the operational logic of various devices. 
    \item We analyze the electrical response characteristics of a quartz crystal oscillator to acoustic vibration both theoretically and experimentally. We observe that the phase and amplitude of the time-based signal output by the crystal oscillator vary for acoustic signals with different phases and amplitudes. Based on the response analyses, we design the principles for modifying the RTC time.
    \item We conduct a set of experiments to assess the security risk posed by TimeTravel on nine off-the-shelf modules and seven commercial devices and propose countermeasures against this security threat.
\end{enumerate}

The rest of the paper is organized as follows. Section 2 describes the background of a quartz crystal oscillator and sound propagation module. Section 3 presents the threat model, including the attack goal, common attack scenarios, and attack assumptions. Section 4 presents the detailed methods to modify the RTC time. Section 5 details the attack steps. Section 6 evaluates the performance of TimeTravel. Section 7 discusses countermeasures, safety recommendations, and the impact of NTP synchronization. Section 8 surveys the related work on acoustic attacks, and finally, Section 9 concludes the work.

\section{Background}
In this section, we present the characteristics and operational logic of a quartz crystal oscillator, as well as the principle of Lamb waves.

\subsection{Quartz Crystal Oscillator}
%\textcolor{red}{
A quartz crystal oscillator is a kind of circuit featured by frequency selection, with a quartz crystal as its principal element\cite{matthys1983crystal}. The quartz crystal oscillator works under \textit{Piezoelectric} effect\cite{katzir2006discovery}, characterized by the crystal's periodic expansion and contraction in response to an applied alternating electric field. Such changes in the crystal's volume generate surface charges to offset the structural changes, thereby generating alternating electrical signals within the circuit.

% The quartz crystal oscillator is a frequency-selective circuit that utilizes a quartz crystal as its core component\cite{matthys1983crystal}. The quartz crystal possesses a significant property called \textit{Piezoelectric effect} \cite{katzir2006discovery}, wherein it undergoes periodic compression or stretching of its volume when subjected to an alternating electric field. The variation in crystal volume induces charges on its surface to counterbalance the deformation, thus forming the alternating signal in the circuit. 

% Quartz crystal oscillator experiences a very fast piezoelectric response speed, typically within a few microseconds or shorter time scales. This enables them to provide high-precision clock and frequency stability in many applications.  Owing to its high quality factor \cite{qfactor} and minimal frequency deviation caused by temperature fluctuations in its surroundings, the quartz crystal oscillator offers superior accuracy and stability compared to other oscillation modules, such as the CMOS oscillator \cite{crystaloscr}. 

%\textcolor{red}{
The quartz crystal oscillator displays an exceptionally high-quality factor \cite{qfactor}, resulting in reduced mechanical energy loss near its resonance frequency. This quality especially enables the crystal responsive to external vibrations close to this frequency, showcasing a swift piezoelectric response, typically in the microsecond range or less\cite{matthys1983crystal}. By contrast, at off-resonance frequencies, the crystal shows increased mechanical energy dissipation\cite{matthys1983crystal}. Therefore, the quartz crystal oscillator is distinguished by its enhanced precision and stability when compared to alternative oscillation devices like CMOS oscillators.
% The quartz crystal oscillator experiences exceptionally high-quality factor \cite{qfactor}, leading to minimal mechanical damping near the resonant frequency. This attribute renders crystal acutely sensitive to external mechanical vibrations that near the resonant frequency, and demonstrates a rapid piezoelectric response, generally occurring within a few microseconds or even shorter timeframes\cite{matthys1983crystal}. Conversely, at frequencies other than the crystal's resonant frequency, it experiences significantly elevated mechanical damping\cite{matthys1983crystal}. Consequently, the quartz crystal oscillator offers superior accuracy and stability compared to other oscillation modules, such as the CMOS oscillator \cite{crystaloscr}. 

% The typical oscillator manufacturing model is shown in Fig. \ref{crysmanufacture_module}. The upper and lower surfaces of the crystal oscillator chip are attached by a pair of metal electrodes, which extend outward from the pins to the external circuit. The entire quartz crystal and electrodes are wrapped by a metal case.

The type of crystal oscillator usually adopted in RTC circuits is called a Tuning-fork-shaped oscillator, and the typical structural model is shown in Fig. \ref{crysmanufacture_module}. The surfaces of the crystal are attached by electrode plating, which extends outward from the pins to the external circuit. The entire quartz crystal and electrodes are wrapped by a metal cover. %\textcolor{red}{
In common oscillator designs, the quartz crystal plays the role of the inductive component. The Pierce oscillator \cite{qfactor} is often preferred for quartz crystal oscillator circuits, as illustrated in Fig. \ref{pierceoscillator_module}. This configuration employs two capacitors, which are connected to the ground and the quartz crystal, respectively, forming a $\pi$-shaped LC frequency-selective network. The inverting amplifier within a microcontroller unit (MCU) is used for providing positive gain to the circuit and introducing additional phase shifts to the signal, ensuring the oscillation signal's stability. With specific configurations of crystal properties and capacitance, the oscillator is designed to have a predetermined resonant frequency. Signals matching this frequency encounter the lowest impedance, facilitating their passage through the crystal\cite{vittoz1988high}.

The RTC circuit typically utilizes a quartz crystal oscillator, resonating at \SI{32.768}{\kilo\hertz}, to generate periodic timing signals. The signal emitted by the oscillator is fed into the counters within the RTC circuit to create a base signal. RTC supports second-level timing (calendar mode), and some RTCs support millisecond-level timing (32-bit mode). For the former, the built-in frequency divider uses a 16-bit counter to divide the crystal oscillator clock pulse into 1 second and output the second-level clock pulses. The latter will be divided into millisecond or microsecond-level clock pulses for accurate timestamp counting. When the rising edge of the oscillation signal exceeds the threshold, the frequency division counter will decrease its stored value by 1, until the counter value drops to 0, and then the counter will trigger a clock pulse and reset. The triggered pulse should be the time-base signal after frequency division.

\subsection{Propagation of Sound in Solid Medium}
Sound is a vibrational disturbance that travels as an acoustic wave through mediums, such as fluids and solids. 
In a solid medium, sound waves can cause both volume deformation (the change in the overall volume of a solid) and tangential deformation (relative displacement or rotation of particles within a solid). In an isotropic medium, the direction of vibration indicated by this shear is perpendicular to the direction of wave propagation\cite{munjal2008formulas}.

Sound waves propagate in different forms inside and on the surface of a solid medium. Inside a solid medium, sound waves mainly propagate as longitudinal waves. By contrast, sound waves propagate as guide waves - surface waves (e.g., Rayleigh waves\cite{rayleighwave}) or Lamb waves on the surface. When the thickness of the solid medium is at least several times larger than the wavelength of the sound wave, surface waves become the dominant form of propagation. Surface acoustic waves attenuate quickly after leaving on one side of the solid surface, usually not exceeding a few wavelengths of depth.
%So, surface waves usually propagate only on one side of the solid surface\cite{munjal2008formulas}.
However, when the solid medium is relatively thin (usually no larger than several times the wavelength), Lamb waves should be the main waveform, and both the top and bottom of the solid surfaces will propagate vibrations\cite{munjal2008formulas}. Since most devices are placed on a table no more than a few centimeters thick, we focus mainly on Lamb waves.

Lamb waves propagate on solid surfaces with two different modes: Symmetric and Anti-symmetric\cite{ling2017review}. These two sets of waves with different modes can propagate independently on the same surface of the solid.  Assuming that the Lamb wave propagates in the $XZ$-plane in the $Z$-direction (see Fig. \ref{propagation_dic}). 
% The displacements caused by the anti-symmetric Lamb wave components to the solid media are mainly clustered in the $X$-direction, while the displacements caused by the symmetric Lamb wave are mainly clustered in the $Z$-direction.
Anti-symmetric Lamb wave components primarily displace the solid media in the $X$-direction, whereas symmetric Lamb waves mainly cause displacements in the $Z$-direction\cite{ling2017review}.

For most kinds of solid medium, driven by low-frequency acoustic excitation signals (usually less than \SI{100}{\kilo\hertz})\cite{hora2012determination}, they will excite the first-order Lamb wave pattern with the same natural angular frequency as those of excitation signals (see Appendix \ref{drlambwavefunc}). Since TimeTravel focuses on the displacement of the medium along the $X$-direction, the equation for the displacement of the medium along the $X$-direction for a specific location $z$ can be expressed as:
\begin{align}
    \xi = \gamma + \lambda \sin(\omega t + \Phi), \label{eq:eq2}
\end{align}
where $\omega$ represents the angular frequency of the excitation sound wave, and $t$ represents the time. The detailed definition of the coefficient $\gamma$, $\lambda$, and $\Phi$ can be found in Appendix \ref{drlambwavefunc}.

\section{Threat Model}

In this section, we introduce the attack goals and assumptions. 

\subsection{Attack Goal}
\begin{figure}[t]
\vspace{-0.1in}
    \centering
    \begin{subfigure}[t]{0.13\textwidth}
        \centering
        \includegraphics[width=\textwidth]{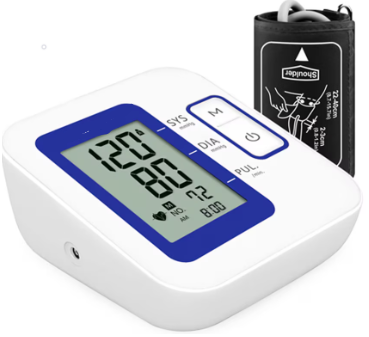}
        \caption{Digital blood pressure monitor.}
    \end{subfigure}
    \hfill % Creates horizontal space between the images
    \begin{subfigure}[t]{0.13\textwidth}
        \centering
        \includegraphics[width=\textwidth]{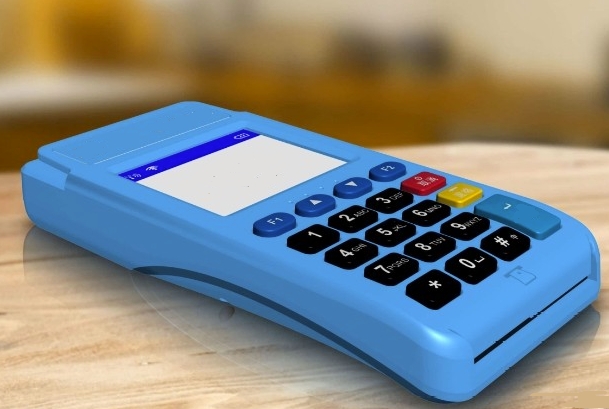}
        \caption{Commercial portable POS machine.}
    \end{subfigure}
    \hfill
    \begin{subfigure}[t]{0.13\textwidth}
        \centering
        \includegraphics[width=\textwidth]{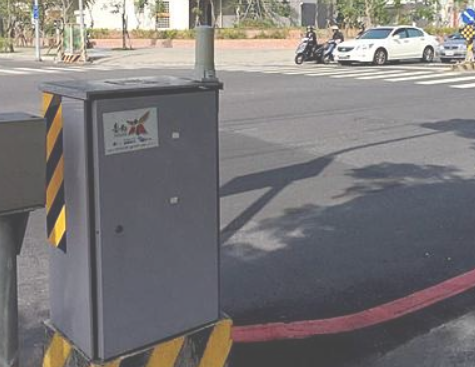}
        \caption{Traffic light controller. }
    \end{subfigure}
    \hfill
    \caption{IoT devices containing real-time clock.}
    \label{fig:scenarios}
    \vspace{-0.1in}
\end{figure}

%The objective of this work is 
This work reveals a serious security risk: time data stored within the RTC circuit can be affected by external acoustic disturbances. Specifically, if a device relies on system time for specific tasks, it may experience a time drift. This drift can be either forward or backward, triggered by well-designed acoustic vibrations. The operational logic and states of these devices can be altered, leading to several potentially severe consequences. Such a vulnerability widely exists in our daily lives. As illustrative examples, we consider real-life scenarios where patients measure and record their basic vital signs, such as blood pressure and heart rate via a digital blood pressure (BP) monitor at a clinic's front desk, before their consultation with doctors (Fig. \ref{fig:scenarios}(a)); customers use POS machines to check out at self-service restaurants or shops (Fig. \ref{fig:scenarios}(b)). All of these devices use the RTC to accomplish specific tasks: the RTC 32-bit mode counters in a BP monitor generate millisecond timestamps to the microcontroller to process the measured pressure signals and control the deflating of the cuff; POS machines use the RTC to read the time and create the records of operations with specific timestamps.

The adversaries' goal is to intentionally emit acoustic wave sequences with specific frequencies for certain purposes. The wave sequence is transmitted to a target device through a solid surface (e.g., a table), and then it can modify the timestamps generated by the device. For example, attackers can maliciously adjust the measurement data of a blood pressure monitor by adjusting internal timestamp, causing the measurement data to be either too high or too low, which will pose a serious health threat even a life threat to a patient as doctors would be misled to take wrong medical treatments to the patient (e.g., issuing anti-hypertensive drugs, but the actual patient's complaint symptoms are not caused by high blood pressure).
%causing patients to experience dizziness, and even shock, seriously threatening the patient's health, and causing a decline in the clinic's reputation; 
For a POS machine scenario, attackers can deliberately adjust the timestamp forward or backward to fall into the promotion time of some commodities, and purchase them with much lower price, resulting in substantial financial gains to themselves but losses to merchants. 
%Those commodities are discounted for a limited time and quantity at these time points. If a specific quantity of discounted commodities is sold out in advance, other customers will not have the opportunity to participate in the rush, causing harm to their interests. 
Furthermore, RTC timestamp also plays a key role in many daily scenarios, like traffic light controlling (Fig. \ref{fig:scenarios}(c)), industrial automation\cite{rtcinplc}, autonomous driving\cite{mozaffari2020deep}, and networking systems\cite{lohrasbinasab2022statistical}. The timing drift in those mission-critical systems will pose serious threats to the society.

\subsection{Attack Assumptions}
We make the following assumptions when mounting the aforementioned attacks. 

\textbf{Attack Requirements.}  TimeTravel is a close-proximity attack; however, attackers only need to modify the time or timestamp of a target device in a controllable way by using acoustic guide waves. 
%Attackers do not need to install malicious software or send network packets (e.g., NTP synchronization packets\cite{ntpdoc}) to the target device. 
Attackers can carefully select the attack parameters to ensure the interference during the attack only affects the crystal oscillator and will not cause any human-perceived abnormalities in the surrounding environment. 

\textbf{Attacker's Prior Knowledge.} We assume that attackers can test the target device on similar mediums under different amplitudes and initial phases of excitation signals, and can verify the corresponding front-end behaviors in advance. This allows attackers to establish the $\varphi \rightarrow \beta_1$ phase mapping (see Section 4.2). Attackers can purchase devices of the same models to conduct the testing.

% We assume that attackers have tested the attack parameters on target devices of the same model in similar environments (e.g., a table made of the same surface material) before launching the attack. Attackers can purchase devices of the same models to conduct the testing.

% or conduct the testing on target devices when nobody is paying attention to.

\textbf{Attack Equipment Setup.} We assume that attackers place a piezoelectric transducer and a magnetic probe in proximity to the target device, 
for emitting guided waves and analyzing the EM signals leaked from the target device, respectively.  In end-to-end attack scenarios, attackers can access the area surrounding the target device (e.g., a table) in advance and position a transducer either in front of or behind the device. A magnetic probe must be strategically placed beneath the device to capture leaked electromagnetic (EM) signals. Attackers can test the target device while it is left unattended in the actual attack environment, to establish the $\varphi \rightarrow \beta_1$ mapping. Based on the mapping and the actual distance between the transducer and the device, attackers select the phase of the oscillating signal detected by the probe. Then they attempt to emit an excitation signal that corresponds to the lowest amplitude of the superimposed oscillating signals. If the phase amplitude of the oscillator decreases and stabilizes during testing, the attack setup is confirmed to be calibrated and ready for use.
To enhance concealment, attackers can attach adhesive tape to the surface of the magnetic probe and transducer, or cover them with camouflage materials.
%attackers can use a USRP and signal amplifier 
To generate and amplify attack signals, attackers need to use a USRP and a signal amplifier, but they can place/hide them far away from the attack plane.

\section{Attack Mechanism Exploration}

In this section, we theoretically explore the injection of attack signals into the RTC circuit for affecting its timing. We then discuss how to design the signal parameters to cause the RTC timestamp to drift forward or backward, followed by simple experimental validations.

\subsection{Principle of Attack Signal Injection}
TimeTravel leverages the sensitivity of the crystal oscillator to resonant frequency by applying additional mechanical vibrations. These vibrations may generate extra electrical signals, thereby altering the characteristics of the original oscillation signal in the circuit, with the intent of influencing the RTC's timing.

To introduce mechanical resonance interference in the crystal oscillator, TimeTravel utilizes a solid medium with the aid of acoustic mechanical displacement, instead of directly emitting ultrasonic waves in the air (see details in Appendix \ref{validsoundpropfalse}). This method propagates emitted mechanical energy into the crystal oscillator. Driven by the vertical displacement of a mass at a specific location on the surface of a solid medium, the rigid object attached to the surface at that location moves vertically with the same acceleration as a whole.
Given that the electrodes are affixed to both ends of the quartz chip and their extended pins are soldered onto the PCB, the quartz crystal oscillator as a whole is thus compelled to undergo the same accelerated mechanical movements as part of the motherboard to which it is attached. When the frequency of the mechanical vibrations applied to the crystal matches its resonant frequency, the crystal induces an additional resonant electrical response. The frequency of this electrical response should be consistent with the frequency of the imported mechanical vibrations (see detailed derivation in Appendix \ref{derimechstress}).

\subsection{Principle of Time Drifting Backward}
% Based on mechanical resonance theory, when a mechanical wave is introduced to a quartz crystal already in resonance, it triggers an additional vibration response.

The influence of resonance response on the crystal's vibration is determined by the phase and amplitude difference between the response of imported vibration and the original resonant signal in the circuit (see Appendix \ref{dersupexpression}). 
To achieve a backward drift of time relative to the current time, the key is to slow down the rate at which the time counter updates as much as possible.
% This can be achieved by minimizing the amplitude of the superimposed oscillation signal, making it not exceed the edge-triggering threshold.
According to Eq. \ref{superimpose_eq}, to achieve this, attackers need to ensure that the phase of the injected 
electrical signals into the oscillator $\beta_1$ and the phase of the original oscillation signal $\beta_2$ differ by as close to $\pi$ as possible. This way, $\cos(\beta_1 - \beta_2)$ approaches -1, and $\phi$ reaches its minimum value. 

As mentioned in Section 4.1, attackers generate mechanical vibrations from a certain distance to the target device, inducing the crystal oscillator to produce additional electrical responses. According to Eq. \ref{eq:eq4}, the initial phase of the electrical response signal $\beta_1$ typically does not match the initial phase of the mechanical acceleration experienced by the crystal oscillator. This discrepancy is affected by the intrinsic properties of the quartz crystal (e.g., damping and stiffness). Additionally, other nonlinear components in the oscillation circuit may introduce a constant phase shift in the signal. These factors make it challenging to directly calculate the initial phase of the electrical response signal $\beta_1$, which is excited by the crystal's mechanical resonance when a specific phase of the electrical signal is applied to the transducer and propagated via mechanical vibration. Instead, we propose an experimental method to determine the mapping relationship between the phase of the excitation signal to the transducer $\varphi$ and the injected electrical response signal $\beta_1$. Attackers can choose the horizontal distance $z$ between the transducer and the crystal, as well as the initial phase $\varphi$ of the excitation signal, and use a magnetic probe placed below the crystal oscillator to record the current oscillation signal phase $\beta_2$ in the circuit, and then immediately emit the signal. The excitation signal will reach the crystal after the propagation delay time $t_T$ (see Eq. \ref{dispeq_xz}), leading to the possible amplitude and phase changes of emanated signals from the crystal oscillator. Attackers can keep $\{z,\varphi\}$ the same and emit signals at different times, after collecting sufficient samples, attackers then obtain values of $z,\varphi$, and $\beta_2$ that result in the relatively lowest observed EM signal amplitude. Referring to Eq. \ref{superimpose_eq}, at this point, $\beta_1-\beta_2$ should approach $\pi$, then attackers can further establish a mapping table $\{z,\varphi\}\rightarrow \beta_1 (\text{note that } \beta_1=\beta_2+\pi)$. Meanwhile, if $\varphi$ changes $x$ (i.e., $\varphi \leftarrow \varphi +x$), $\beta_1$ will change $x$ accordingly.

In the subsequent attack, attackers can select the phase $\varphi$ and the attack distance $z$, mapping it to $\beta_1$. After observing that the phase of the signal emitted by the oscillation circuit reaches $\beta_2=\beta_1-\pi$, attackers immediately transmit and maintain the attack signal, so that the signal amplitude outputted from the oscillator will remain stable at a relatively low value after interference. The value $k_a$ in Eq. \ref{eq:eq2} can be determined by solving  Eq. \ref{eqset1}\cite{rose2000ultrasonic}, which refers to the characteristic equation of Lamb waves:
\begin{equation}
 \frac{\tan(\sqrt{\frac{\omega^2}{c_L^2}-k_a^2}d)}{\tan(\sqrt{\frac{\omega^2}{c_T^2}-k_a^2}d)}=-\frac{4\sqrt{(\frac{\omega^2}{c_T^2}-k_a^2)(\frac{\omega^2}{c_L^2}-k_a^2)}k_a^2}{(\frac{\omega^2}{c_L^2}-2k_a^2)^2},
\label{eqset1}
\end{equation}
where the angular frequency $\omega$ = $2\pi f$, $d$ represents the thickness of the solid medium, and $c_L$ and $c_T$ represent the velocities of longitudinal and transverse waves inside the medium, respectively, which can be found in Table \ref{mediumdata}. 

\begin{figure}[t]
    \centering
    \begin{subfigure}[t]{0.16\textwidth}
        \centering
        \includegraphics[width=\textwidth]{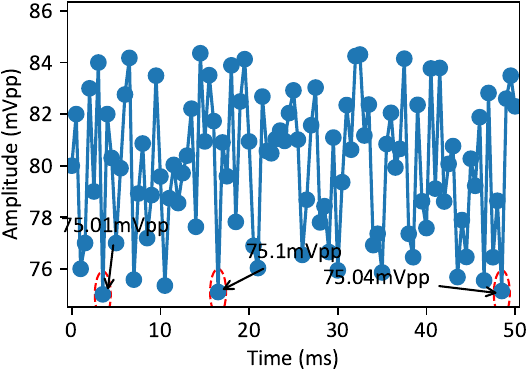}
        \caption{}
    \end{subfigure}
    \hfill % Creates horizontal space between the images
    % \begin{subfigure}[t]{0.22\textwidth}
    %     \centering
    %     \includegraphics[width=\textwidth]{pics_and_tables/pic_7.pdf}
    %     \caption{}
    % \end{subfigure}
    % \hfill
    \begin{subfigure}[t]{0.16\textwidth}
        \centering
        \includegraphics[width=\textwidth]{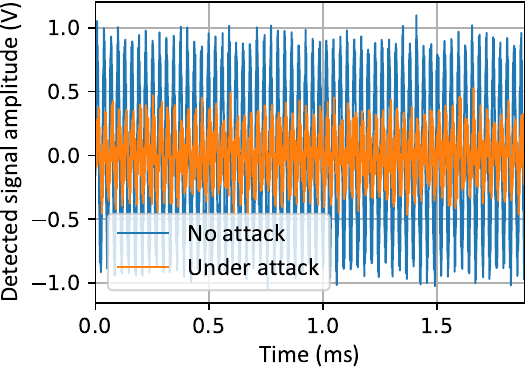}
        \caption{}
    \end{subfigure}
    \hfill
    \begin{subfigure}[t]{0.13\textwidth}
        \centering
        \includegraphics[width=\textwidth]{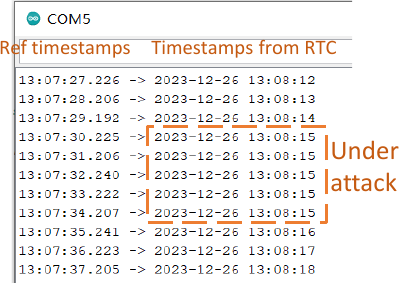}
        \caption{}
    \end{subfigure}
    \hfill
    \caption{Experimental results on the backward drift of time.}
    \label{fig:combined}
\end{figure}

To verify the feasibility of time drifting backward, we conduct a sample experiment by employing a USRP N210 equipped with an amplifier to generate sinusoidal signals at a frequency of \SI{32.768}{\kilo\hertz}, an initial phase $\varphi=0$ and an amplitude of 20Vpp. Other experimental apparatus includes an acrylic glass plate with a \SI{5}{\mm} thickness and a piezoelectric ceramic transducer\cite{piezodisc} with \SI{3}{\cm} in diameter and \SI{2}{\mm} in thickness. This transducer is connected to the amplifier, facilitating the transmission of Lamb waves to the glass plate. We integrate a DS1302 RTC module \cite{ds1302mod} with an Arduino Uno Rev3 board using DuPont wires for further experimentation. The ceramic transducer is positioned roughly \SI{5.5}{\cm} from the RTC module's quartz oscillator, with a magnetic field probe placed directly behind the glass plate beneath the module, as depicted in Fig. \ref{propagation_solid}. We emit pulses of 0.3ms duration at 0.5ms intervals, measuring the signal amplitude at the crystal oscillator both before and during these emissions over a 50ms period. We observe that the detected amplitudes of oscillating signals, as presented in Fig. \ref{fig:combined}(a), have significant reductions at \SI{3.2}{\ms}, \SI{16.3}{\ms}, and \SI{48}{\ms}, with recorded values of 75.01mVpp, 75.1mVpp, and 75.04mVpp, respectively; the phase of oscillating signals $\beta_2$ at these timestamps is around 0.56 deg. These values represent approximately 93.8\% of the baseline amplitude of 80mVpp, suggesting that the initial phase of the electrical signal injected into the crystal oscillator should be $\beta_1=(3.14+0.56)$ deg. We further observe that the peak-to-peak amplitude of the superimposed signals remains consistently low during the attack. We increase the amplitude of the excitation signal and observe that as the amplitude increases, the detected oscillation signal amplitude gradually decreases. When the applied signal amplitude reaches 65Vpp, we notice that the oscillation signal's maximum amplitude decreases and then stabilizes at 36mV (Fig. \ref{fig:combined}(b)). Additionally, we observe that the RTC module's timing stops (Fig. \ref{fig:combined}(c)), but it resumes to normal once the attack ceases.

\subsection{Principle of Time Drifting Forward}
To expedite clock timing, attackers should design attack signal sequences to alter the phase of oscillation signals, thereby ensuring that the amplitude of oscillation signals exceeds the edge-triggering threshold more frequently. Assume that the initial phase of the electrical interference signal caused by the acoustic vibration is $\beta_1$, and the phase of the oscillation signal in the oscillator is $\beta_2$. We denote $\beta_1$ as $\beta_1=\beta_2+\Delta $. According to Eq. \ref{superimpose_eq}, the phase of the superimposed signal can be expressed as:
 \begin{align}
      \beta_2'&=\frac{\beta_1+\beta_2}{2}+\arctan(\frac{A-B}{A+B}\cdot \tan(\frac{\beta_1-\beta_2}{2})) \notag\\
         &=\beta_2+\frac{\Delta}{2}+\arctan(\frac{A-B}{A+B}\cdot\tan(\frac{\Delta}{2})),
     \label{eq:eqphase}
     \end{align}
where $A,B$ refer to the amplitude of attack electrical signal and oscillation signal, respectively. The superimposed oscillation signal will have a new phase $\beta_2' (0<\beta_1-\beta_2'< \Delta)$, and will be immediately amplified by the feedback network and then sent back to the quartz crystal. The oscillation signal with phase $\beta_2'$ will further overlap with the interference signal applied by attackers. After a period of time, the phase difference between the electric resonance signal of the crystal oscillator and the attack signal will gradually decrease until the phase remains consistent. Finally, the phase of the oscillation signal will be $\Delta$ earlier than before being attacked. Afterward, attackers can continue to emit attack signals with a phase difference of $\Delta$ from the current oscillation signal, causing the phase in the oscillation signal to continuously drift forward. 

Using the superposition of signals of the same frequency to change the phase is equivalent to phase modulation, which broadens the signal spectrum in the frequency domain\cite{roder1931amplitude}.
When the attack lasts for a period of time (usually not exceeding a dozen microseconds), the frequency band around the \SI{32.768}{\kilo\hertz} should gradually narrow and coincide with the central frequency. 
Then, it can be assumed that the phase of the oscillation signal in the circuit converges to the injected attack signal. Attackers can observe the phase of the oscillation signal at an appropriate time based on the phase mapping $\{z,\varphi\}\rightarrow \beta_1$ obtained in Section 4.2, causing $\beta_2$ to move $\Delta$ forward. After the phase of the oscillation signal converges and stabilizes, attackers can continue to shift the attack signal phase forward by $\Delta$, to induce $\beta_2$ to shift forward by $\Delta$. After each attack, attackers will force the next rising/falling edge level trigger time to arrive $\frac{\Delta}{2\pi}$ earlier, as shown in Fig. \ref{samprotation1}. Considering that the interval between the rising and falling edge trigger points equates to half an oscillation cycle,  attackers can establish $0<\Delta<\pi$ to ensure the integrity of the edge triggering.

\begin{figure}[t]
\vspace{-0.1in}
    \centering
    \begin{subfigure}[t]{0.42\textwidth}
        \centering
        \includegraphics[width=\textwidth]{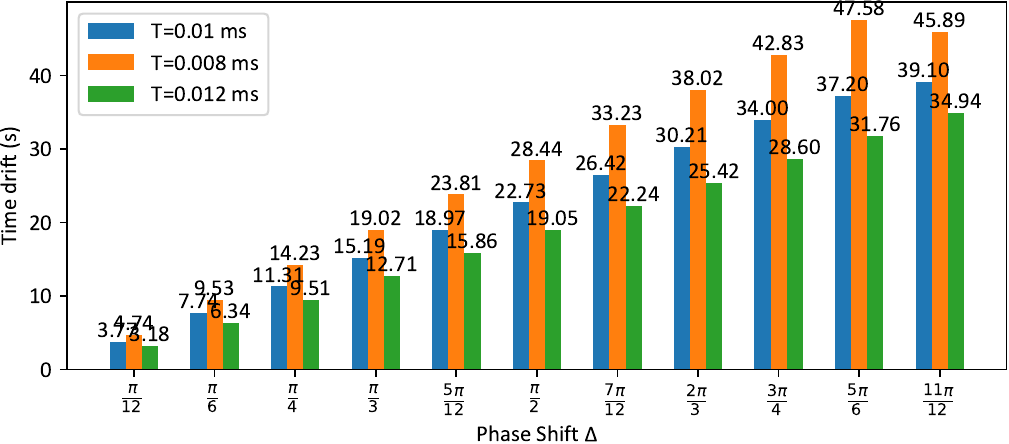}
        \caption{}
    \end{subfigure}
    \hfill % Creates horizontal space between the images
    % \begin{subfigure}[t]{0.2\textwidth}
    %     \centering
    %     \includegraphics[width=\textwidth]{pics_and_tables/pic_12.pdf}
    %     \caption{$\Delta=\frac{\pi}{12}$ (blue line) and $\frac{11\pi}{12}$(green line).}
    % \end{subfigure}
    % \hfill
    % \begin{subfigure}[t]{0.2\textwidth}
    %     \centering
    %     \includegraphics[width=\textwidth]{pics_and_tables/pic_13.pdf}
    %     \caption{}
    % \end{subfigure}
    % \hfill
    \caption{Experimental results on forward drift of time.}
    \label{timeforwards_res}
    \vspace{-0.2in}
\end{figure}

To validate the feasibility of time drifting forward, we use a \SI{5}{\mm}-thick acrylic glass plate and place the transducer \SI{3}{\cm} next to the DS1302 RTC crystal, and then when we observe that the oscillation signal emitted by the oscillator has a phase of $\frac{\pi}{2}$, we begin to emit excitation signals of \SI{32.768}{\kilo\hertz} that cause the oscillation signal phase to shift forward $\Delta$. After every $T$ ms of attack, attackers shift the phase of the excitation signal forward by $\Delta$. We set the amplitude of the excitation signal to 65Vpp, the phase shift offset $\Delta=\frac{n\pi}{12} (n=1,2,...,11)$ , and $T=0.008,0.01,0.012$ ms. For each phase offset, we emit attack signal sequences \SI{30}{\second} and repeat the test 50 times. For each selected phase offset, we record the average offset of the time change read by RTC relative to the reference time change, as shown in Fig. \ref{timeforwards_res}(a). 

Overall, the time drift of RTC increases with the increase of phase drift in a single attack. When the phase drift is $\frac{\pi}{12}$, the cumulative time drift under the three attack durations does not exceed \SI{5}{\second}. When the phase drift is $\frac{11\pi}{12}$, the timing rate of RTC is at least doubled. Meanwhile, the duration of the excitation signal emission also affects the final time drift. When the duration of signal transmission attacks is shortened, the overall time drift will increase accordingly. Under the phase shift $\Delta=\frac{11\pi}{12}$, the change in attack time from \SI{0.012}{\ms} to \SI{0.008}{\ms} will result in an additional time drift of approximately \SI{13.75}{\second}. This is because when the duration of the attack signal is shortened, the phase of the oscillation signal will shift forward more frequently, causing the oscillation signal to cross the edge trigger level more frequently. Under relatively long attack signal duration, the phase of the oscillating signal will change synchronously with the attack signal after converging to be consistent with the attack signal, without causing any additional phase shift and time forward drift. 
However, if the attack duration is too short ($T=0.008$ ms), we observe that under a large phase drift $\frac{11\pi}{12}$, the crystal does not fully respond to the vibration mode of the attack signal, resulting in the oscillation signal bandwidth still not approaching 0 at the end of a single attack, that is, the oscillation signal phase does not converge to the attack signal. Therefore, time does not drift forward steadily at a specific rate, resulting in an unexpected increase in time drift.

\section{Attack Design}

\begin{figure}[t]
%\vspace{-0.1in}
\centering
\centerline{\includegraphics[width=0.4\textwidth]{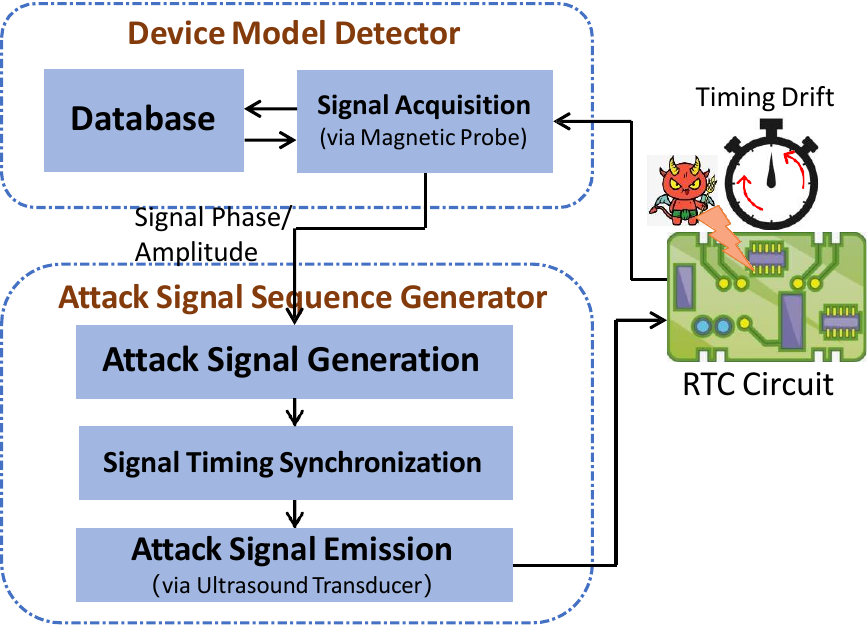}}
\caption{The general design of the TimeTravel attack scheme.}
\label{attackscheme_pic}
%\vspace{-0.1in}
\end{figure}

After exploring the feasibility of transmitting sound waves through acoustic transducers to affect the accuracy of clock timing within RTC, we now present the design principles for TimeTravel to achieve controllable and quantitative time drift attacks. To mount a successful attack, attackers should address the following two key challenges: (1) How to identify device models to match appropriate attack parameters? (2) How to configure the parameters of the attack signal (e.g., attack distance, phase, interval, and amplitude) to achieve the desired timing rate? Specifically, we elaborate on the design of the TimeTravel attack scheme, including the methods for identifying the model of devices, as well as attack signal configuration, as shown in Fig. \ref{attackscheme_pic}.

\subsection{Detection of the Target Device's Model}
Different devices typically use distinct PCB materials, layouts, and RTC crystal oscillator models, resulting in varied stress distributions and vibration response characteristics under mechanical vibration. Detecting the model of the target device allows attackers to choose an appropriate attack strategy. One method is to leverage the frequency uniqueness of MCU clock signal components to recognize device models.

% Different types of devices usually adopt different PCB materials, layouts, and RTC crystal oscillator models. This causes different stress distributions when the RTC circuit is subjected to mechanical vibration, resulting in different vibration response characteristics. The detection of the target device's model enables attackers to decide the appropriate attack strategy.
% \textcolor{red}{One possible method is to leverage the frequency uniqueness of MCU clock signals to recognize device models.} Different models of devices typically adopt different MCUs and surrounding electrical components.
% Attackers can use a magnetic probe to collect electromagnetic signals leaked from the device's MCU over a fixed period, and then use neural networks (e.g., CNN\cite{albawi2017understanding}) to extract and classify the characteristics of the electromagnetic signals, obtaining the probability distributions belonging to different types of MCUs. Then, attackers can query the device model corresponding to the detected MCU in the database and obtain the attack parameters that have been previously recorded.
Attackers can use a magnetic probe to collect electromagnetic signals leaked near the device's MCU chip over a fixed period. After acquiring the raw EM signals, TimeTravel scales and normalizes the signal to balance the impact of detection distance on signal strength. Assume that the original signal sequence is $x=\{x_0,x_1,...,x_n\}$, TimeTravel maps the original signal amplitude to $-a$ to $b$ mV ($a,b>0$) through a linear transformation according to Eq. \ref{scaledx}:
\begin{align}
x_{scaled}=-a+(\frac{x-x_{min}}{x_{max}-x_{min}}\cdot (b+a)),
\label{scaledx}
\end{align}
where $x$ represents the original sample value, and $x_{min}$ and $x_{max}$ represent the minimum and maximum values in the original sample data, respectively.

\begin{figure}[t]
%\vspace{-0.15in}
\centering
\centerline{\includegraphics[width=0.26\textwidth]{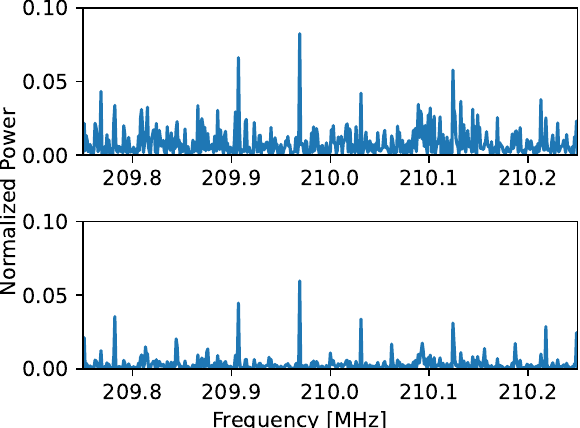}}
\caption{Undenoised signal FFT result (top) and denoised signal FFT result (bottom).}
\label{FFT_1}
%\vspace{-0.1in}
\end{figure}

Manufacturers typically use spread spectrum clocks (SSC) to generate the MCU clock signal, distributing the clock signal energy over a unique frequency band for electromagnetic compatibility. The SSC signal can be expressed as $V(t)=A\sin(2\pi f_0 t+\Delta f \frac{f_0}{f_m} \sin(2\pi f_m t))$, where $f_0$ is the center clock frequency, and $f_m$ and $\Delta f$ are the modulation frequency and frequency offset, respectively. These parameters are determined by the MCU's hardware configuration and manufacturing process differences. Attackers can investigate the MTU clock center frequencies $f_0$ of the target device in advance and add them to the collection $G_0$. After capturing a fixed length of the electromagnetic signal, TimeTravel iteratively selects $f_0$ in $G_0$ as the center frequency and uses a band-pass filter with $f_0$ as the center frequency and a preset bandwidth $M$ to reduce the overall signal bandwidth. Furthermore, TimeTravel uses wavelet denoising \cite{pan1999two} to filter out noise unrelated to the spread spectrum clock signals. For instance, the denoised probing signal from the core MCU of the Omron HEM-7132, as shown in Fig. \ref{FFT_1}, demonstrates that after normalization and denoising, the detection of clock frequency components in the distance of 2 cm from the device's bottom can be enhanced with higher SNR.

To further identify subtle differences in $f_0, f_m,$ and $\Delta f$ among different device models, we use a CNN to classify one-dimensional signal sequences. The network consists of alternating convolution and pooling layers, followed by a softmax layer that classifies the input signal fragments into the device models in the database and outputs the classification confidence for each model. The detailed structure of the model is shown in Table \ref{classifynetwk}. If the confidence level for all models is low in the output layer, it indicates that the target device model is not in the attacker's database, and TimeTravel will not proceed with the classification step.

\subsection{Attack Acoustic Sequence Generation}
Once attackers determine the model of the target device, they can design attack signal sequences to drift the target device's system time/timestamp at a certain rate. For time drifting forward and backward, there are different signal sequence generation strategies.

\subsubsection{Time Drifting Backwards}
The backward drift of time is achieved by utilizing the static logic characteristics of the RTC time counter. However, some RTC time counters may freeze if they do not detect an edge trigger for an extended period, remaining frozen until the device is restarted. To prevent time freezing, it is necessary to distribute the expected time drift length as evenly as possible throughout the entire attack period for accumulation. 
% To avoid the RTC not being able to restore the timing due to the loss of the clock signal for a long time, attackers should limit the duration of a single round of signal emission. 
Assume that attackers expect the internal time in RTC to drift backward by $b$ time units within $a$ time units ($a>b$). If $b$ exceeds the time threshold that causes RTC to freeze, they can design the sequence of attacks by continuously emitting attack signals for $t$, then stopping the attack for $\frac{a-b}{\frac{b}{t}-1}$. Here $t$ should be less than the maximum attack duration that causes RTC to freeze. Attackers should repeatedly send the attack sequence $\lceil \frac{b}{t} \rceil$ times so that the time drift of $b$ time units is evenly distributed across the attack duration of $a$ time units.

% Based on the horizontal distance between the transducer and the RTC crystal, attackers should choose the initial phase of the observed oscillation signal, and design the amplitude and initial phase of the excitation signal based on the phase mapping relationship obtained from the database, to force the amplitude of the superimposed oscillation signal to fall below the threshold that causes the RTC to stop timing.
After launching an attack, attackers need to calculate the phase of the oscillation signal after $t+\frac{a-b}{\frac{b}{t}-1}$ units, and then send an excitation signal corresponding to the attack phase at $t+\frac{a-b}{\frac{b}{t}-1}-t_T$, where $t_T$ represents the time required for Lamb waves to propagate from the transducer to the crystal oscillator, as mentioned in Eq. \ref{dispeq_xz}.

% At the moment of launching each round of the attack signal, the attacker needs to detect the appearance of this phase using a probe and then launch the attack immediately. Due to the frequency of the oscillation signal being \SI{32.768}{\kilo\hertz} (with a period of approximately \SI{0.03}{\ms}), the time spent waiting for targeted phase arrival at the beginning of a round of attack is mostly negligible.

\subsubsection{Time Drifting Forward}
The forward drift of time is achieved by accelerating the time interval of the oscillation signal reaching the edge trigger threshold. 
% Attackers can extract phase mapping relationships $\{z,\varphi \} \rightarrow \beta_1$ from the database; then based on the phase of the observed oscillation signal at the beginning of the attack, attackers generate excitation signals with the initial phase $\varphi$ to transducer that causes the oscillation signal phase to drift forward $\Delta$. 

Assuming that attackers expect the time to drift forward by $b$ time units within $a$ time units, under the maximum time drift range allowed by the attack parameters. 
% To evenly distribute the time drift Assume that attackers expect the internal time in RTC to drift backward the entire attack process, 
To evenly distribute the time drift throughout the entire attack, attackers can choose the duration $t_1 > 0$ of a single attack signal transmission and the interval $t_2 > 0 $ between two consecutive attack signals. Throughout the entire attack, attackers need to emit $k$ attack signals, where $k$ and $t_2$ satisfy:
\begin{equation}
\left\{
\begin{aligned}
    k&=\lceil \frac{2\pi b}{\Delta} \rceil\\
    t_2 &= \frac{a-t_1\lceil \frac{2\pi b}{\Delta} \rceil}{\lceil \frac{2\pi b}{\Delta} \rceil -1},
\end{aligned}
\right.
\label{e1}
\end{equation}
where $\Delta$ refers to the amount of the oscillation signal phase to drift forward. Due to $t_2>0$, the time drift assumed by attackers should satisfy $a> \lceil\frac{2\pi b}{\Delta} \rceil t_1$.
% \begin{equation}
%     a> \lceil\frac{2\pi b}{\Delta} \rceil t_1
% \end{equation}

\begin{table*}[t]
\caption{The Success Rates under Different Modules/Boards with RTC and Parameter Configurations.}
\vspace{-0.15in}
\begin{center}
\begin{tabular}{|c|c|c|c|c|c|c|c|c|c|c|c|}
\hline
\textbf{\#}&\textbf{Module}&\textbf{Pw.$\uparrow$}&\textbf{Pw.$\downarrow$}&\textbf{$\downarrow$5s}&\textbf{$\downarrow$25s}&\textbf{$\downarrow$5.5s}&\textbf{$\downarrow$20.05s}&\textbf{$\uparrow$25s}&\textbf{$\uparrow$45s}&\textbf{$\uparrow$20.5s}&\textbf{$\uparrow$40.05s}\\
\hline
\multirow{2}*{1}&\multirow{2}*{DS1302}&\multirow{2}*{68.2V}&\multirow{2}*{75V}&89\%/&87\%/&\multirow{2}*{---}&\multirow{2}*{---}&88\%/&90\%/&\multirow{2}*{---}&\multirow{2}*{---}\\
&&&&2.519$^{\mathrm{1}}$&2.519&&&1.009&1.979&&\\
\hline
\multirow{2}*{2}&\multirow{2}*{DS1307}&\multirow{2}*{67.1V}&\multirow{2}*{76V}&88\%/&90\%/&\multirow{2}*{---}&\multirow{2}*{---}&88\%/&85\%/&\multirow{2}*{---}&\multirow{2}*{---}\\
&&&&2.515&2.515&&&1.005&1.975&&\\
\hline
\multirow{2}*{3}&\multirow{2}*{PCF8563T}&\multirow{2}*{66.9V}&\multirow{2}*{73V}&89\%/&85\%/&\multirow{2}*{---}&\multirow{2}*{---}&91\%/&86\%/&\multirow{2}*{---}&\multirow{2}*{---}\\
&&&&2.521&2.521&&&1.011&1.981&&\\
\hline
\multirow{2}*{4}&\multirow{2}*{DS3231}&\multirow{2}*{64.6V}&\multirow{2}*{75V}&91\%/&89\%/&\multirow{2}*{---}&\multirow{2}*{---}&88\%/&87\%/&\multirow{2}*{---}&\multirow{2}*{---}\\
&&&&2.527&2.527&&&1.017&1.987&&\\
\hline
\multirow{2}*{5}&STM STM32-&\multirow{2}*{71.8V}&\multirow{2}*{79V}&\textbf{93\%}/&89\%/&87\%/&86\%/&87\%/&86\%/&85\%/&87\%/\\
&F103ZET6&&&2.535&2.535&2.535&2.535&1.025&1.995&0.844&1.922\\
\hline
\multirow{2}*{6}&Alientek XC6-&\multirow{2}*{71.1V}&\multirow{2}*{95V}&84\%/&85\%/&85\%/&86\%/&84\%/&85\%/&84\%/&82\%/\\
 &C6SLX16&&&2.518&2.518&2.518&2.518&2.79&2.395&0.827&1.905\\
\hline
\multirow{2}*{7}&Alientek STM32-&\multirow{2}*{72.3V}&\multirow{2}*{94V}&85\%/&\textbf{83\%}/&86\%/&84\%/&84\%/&84\%/&\textbf{81\%}/&\textbf{78\%}/\\
 &F103ZGT6&&&2.519&2.519&2.519&2.519&2.791&2.399&0.828&1.906\\
 % \cline{1-14}
 \hline
\multirow{2}*{8}&Alientek STM32-&\multirow{2}*{71.9V}&\multirow{2}*{86V}&89\%/&88\%/&90\%/&85\%/&85\%/&87\%/&86\%/&81\%/\\
 &F407ZGT6&&&2.536&2.536&2.536&2.536&2.809&2.413&0.845&1.923\\
  % \cline{1-14}
 \hline
\multirow{2}*{9}&DINGCHANG&\multirow{2}*{73.2V}&\multirow{2}*{88V}&89\%/&91\%/&\textbf{92\%}/&89\%/&\textbf{83\%}/&85\%/&89\%/&\textbf{90\%}/\\
 &DC-A566&&&2.921&2.921&2.921&2.921&2.801&2.396&0.83&1.908\\
  % \cline{1-14}
 \hline
\multicolumn{12}{l}{$^{\mathrm{*}}$ The initial phase of the excitation signal applied by attackers to the transducer, expressed in radian (rad); Pw.$\uparrow$ and Pw. $\downarrow$}\\
\multicolumn{12}{l}{refer to the minimum power transmitted to transducer, to cause the desired forward and backward timing drift, respectively.} \\
\end{tabular}
\label{moduletable}
\end{center}
\vspace{-0.25in}
\end{table*}

\begin{figure}[t]
\vspace{-0.1in}
\centering
\centerline{\includegraphics[width=0.3\textwidth]{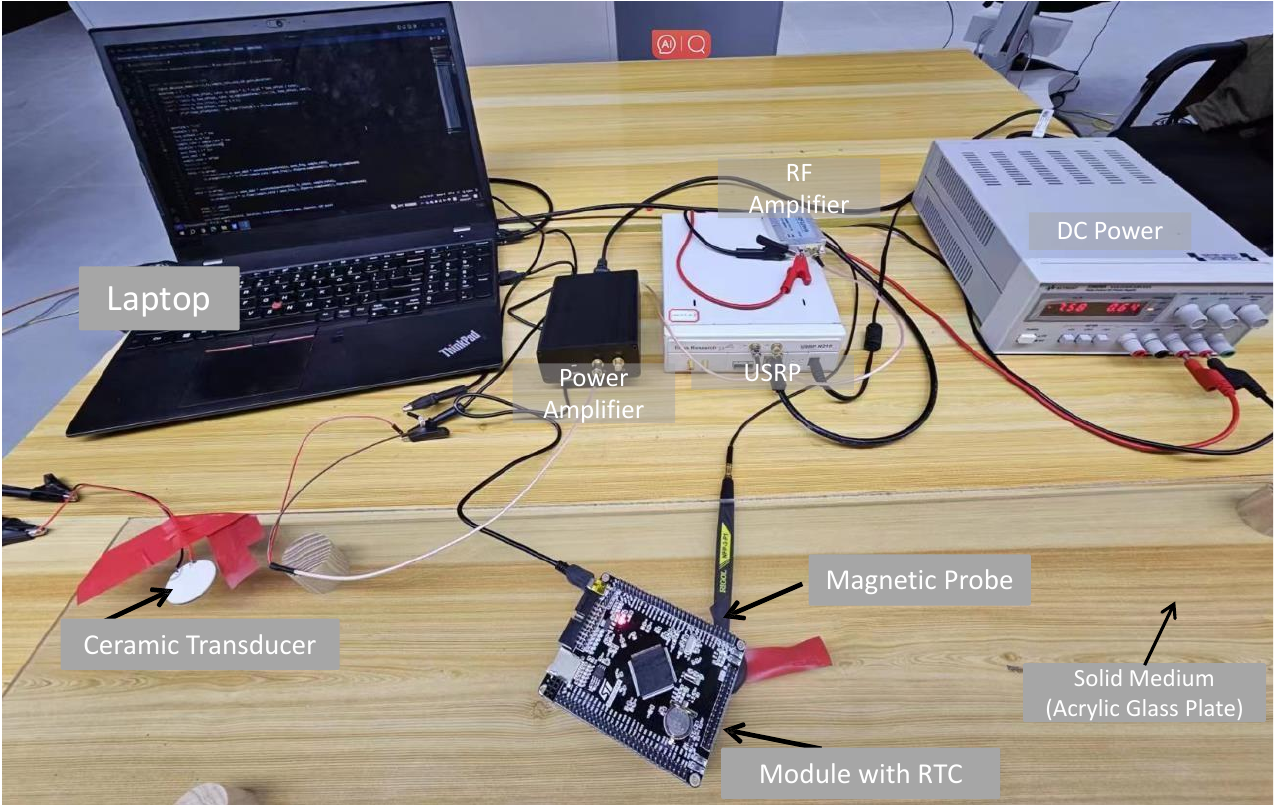}}
\caption{Experimental setup of the TimeTravel evaluation.}
\label{exp_Setup}
\vspace{-0.1in}
\end{figure}

\section{Experiment}
In this section, we first elaborate on our experimental setup and then evaluate the overall performance of TimeTravel.

\subsection{Experimental Setup}
\subsubsection{Attack Device}
We use a RIGOL NFP-3 magnetic field probe set to sense the magnetic field emanating from the crystal oscillator. The probe is connected to an RF probe amplifier (frequency from \SI{10}{\kilo\hertz} to \SI{1}{\giga\hertz}) to increase the amplitude of the sensing signal and filter noise. The received signal is transferred to a laptop by ETTUS USRP N210  for signal analysis and synchronization. For Lamb wave transmission, we adopt a Taimi ultrasonic piezoelectric ceramic transducer (resonance frequency band $15\sim35$kHz, maximum supporting voltage 240Vpp) and connect to FPA101A power amplifier as well as USRP to emit the Lamb wave acoustic signals.

\subsubsection{Device Under Testing}
\label{tenv}
We use four standalone RTC timing modules to test the attack performance of TimeTravel without any irrelevant circuit components, and five types of ARM development boards to test the attack on the RTC circuit under a general IC layout configuration. These boards/modules are listed in Table \ref{moduletable}. For testing standalone RTC modules, we use the Arduino UNO R3 development board, connecting it to the modules to send commands and analyze returned time responses. The experimental setup is shown in Fig. \ref{exp_Setup}. To verify in real scenarios, we test one commercial POS machine (Kaidianbao POS) and six commercial electronic blood pressure monitors (Omron HEM-7132, HEM-7124, HEM-7121; Xiaomi MI-BPX1; HYNAUT AXD-808; Yuwell YE660D). For the testing environment, we use aluminum, wood, acrylic glass, and hard plastic plates of different thicknesses as the solid medium for sound propagation. We set up the layout of our testing environment into two scenarios: In the first scenario, we position the module subjected to the attack on the surface of the solid medium. In the second scenario, besides placing the module on the medium's surface, we also place extraneous items around the module (see Fig. \ref{ac_scenarios_setting}). The second scenario is specifically designed to assess the robustness of TimeTravel.

\subsubsection{Evaluating Metrics}
%We evaluate the performance of TimeTravel under 
Our evaluation is based on the following metrics.
(1)
\textbf{Success rate} is defined as the average probability of reaching the expected time drift within a specified time, under different configuration conditions.
(2)
\textbf{Processing delay} is the time interval when the target phase signal is captured from the oscillator circuit to the beginning of the emission of the acoustic attack signal from USRP.
(3)
\textbf{Robustness} assesses the ability to maintain a successful attack when there are some unrelated objects on the surface of the solid medium, where the attack target is located.

% \begin{enumerate}
%     \item \textbf{Success rate}. The success rate can be expressed as the average probability of reaching the expected time drift within a specified time under different configuration conditions.
%     \item \textbf{Latency}. 
%     \item \textbf{Attack distance}. 
%     \item \textbf{Robustness}. 
% \end{enumerate}

\subsection{Performance Evaluation}
In the experiment, we send commands to ARM/Arduino development boards to record the time from RTC counters. During the testing, we continuously adjust the vertical distance between the magnetic field probe and the oscillator, as well as the amplitude of the excitation signal to the transducer, to determine the maximum effective distance for detecting the phase of the oscillation signal and the minimum signal energy required to cause all desired time drift speeds. For the analysis on commercial devices, due to the inability to directly access the RTC time data, we evaluate the impact of TimeTravel on RTC through the relevant data output from the device (e.g., blood pressure, pulse, and transaction time).

% \begin{figure}[t]
%     \centering
%     \begin{subfigure}[t]{0.08\textwidth}
%         \centering
%         \includegraphics[width=\textwidth]{pics_and_tables/P1_meter.png}
%         \caption{}
%     \end{subfigure}
%     \hfill % Creates horizontal space between the images
%     \begin{subfigure}[t]{0.08\textwidth}
%         \centering
%         \includegraphics[width=\textwidth]{pics_and_tables/Fig2222.png}
%         \caption{}
%     \end{subfigure}
%     \hfill
%     \begin{subfigure}[t]{0.08\textwidth}
%         \centering
%         \includegraphics[width=\textwidth]{pics_and_tables/P2_meter.png}
%         \caption{}
%     \end{subfigure}
%     \hfill
%     \begin{subfigure}[t]{0.08\textwidth}
%         \centering
%         \includegraphics[width=\textwidth]{pics_and_tables/aa.png}
%         \caption{}
%     \end{subfigure}
%     \hfill
%         \begin{subfigure}[t]{0.08\textwidth}
%         \centering
%         \includegraphics[width=\textwidth]{pics_and_tables/f13.png}
%         \caption{}
%     \end{subfigure}
%     \hfill
%     % \begin{subfigure}[t]{0.15\textwidth}
%     %     \centering
%     %     \includegraphics[width=\textwidth]{pics_and_tables/p3_pos.png}
%     %     \caption{}
%     % \end{subfigure}
%     % \hfill
%     \caption{\textcolor{red}{Experimental results under HEM-7132.}}
%     \label{commercial_devices}
%     \vspace{-0.1in}
% \end{figure}

 \subsubsection{Success Rate}
 \label{section_succrate}
\textbf{A. General Attack Accuracy.} 
(1) Evaluation on RTC modules/development boards: We utilize a \SI{5}{\mm}-thick acrylic glass plate as a medium for the experimental setup. The transducer is placed on the top surface of the plate, about \SI{20}{\cm} horizontally from the RTC crystal. We place a magnetic probe on the reverse side of the plate, aligning it with the location of the crystal. We conduct over a 30-second attack for each test and set the single attack duration of \SI{0.009}{\ms} for time forward drift scenarios. 
%During the tests, 
We carefully calibrate the amplitude and initial phase of the excitation signal to align with the expected time drift. For each module under test, 
we standardize the phase of the oscillating signal, observed at the onset of the attack, to zero. Table \ref{moduletable} shows the efficacy of attacks across nine modules, delineating different attack parameters and expected time drift scenarios. The results demonstrate that TimeTravel can effectively induce time drift in RTC across all modules, albeit at different drift speeds. Specifically, the success rate of mounting time modification attacks at a coarse granularity of \SI{1}{\second} varies from \SI{83}{\percent} to \SI{93}{\percent}. For a granularity of \SI{0.1}{\second}, this rate ranges between \SI{81}{\percent} and \SI{92}{\percent}, and for \SI{0.01}{\second}, it ranges from \SI{78}{\percent} to \SI{90}{\percent}. Note that since the selected standalone RTC modules lack the capability of outputting the time at the fine-grained millisecond level, we exclude these modules from the attack tests at the \SI{0.1}{\second} and \SI{0.01}{\second} granularities. The results indicate that no obvious correlation exists between the attack complexity and the precision of the desired time drift.

\begin{figure}[t]
\vspace{-0.1in}
    \centering
    \begin{subfigure}[t]{0.23\textwidth}
        \centering
        \includegraphics[width=\textwidth]{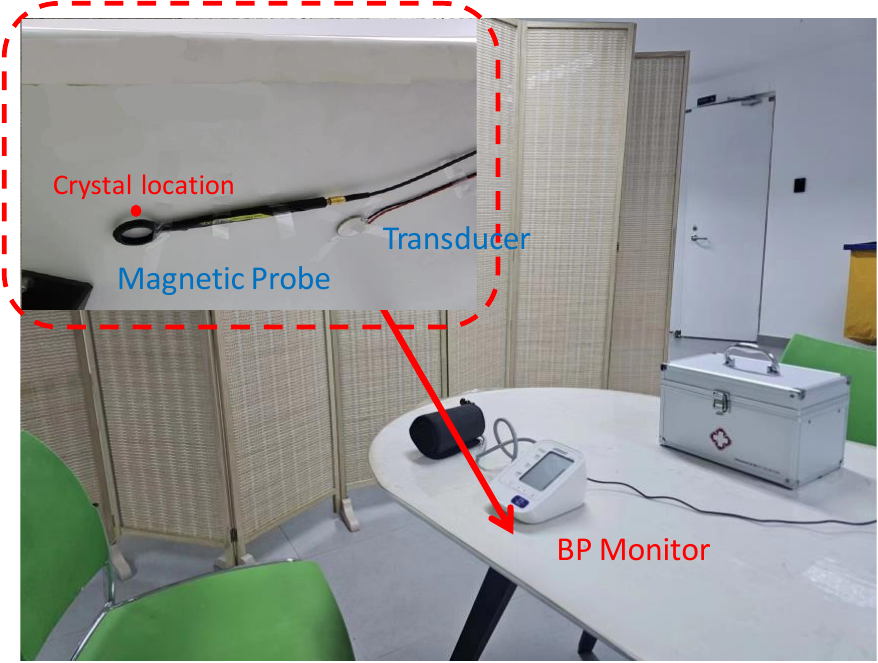}
        \caption{}
    \end{subfigure}
    \hfill % Creates horizontal space between the images
    \begin{subfigure}[t]{0.23\textwidth}
        \centering
        \includegraphics[width=\textwidth]{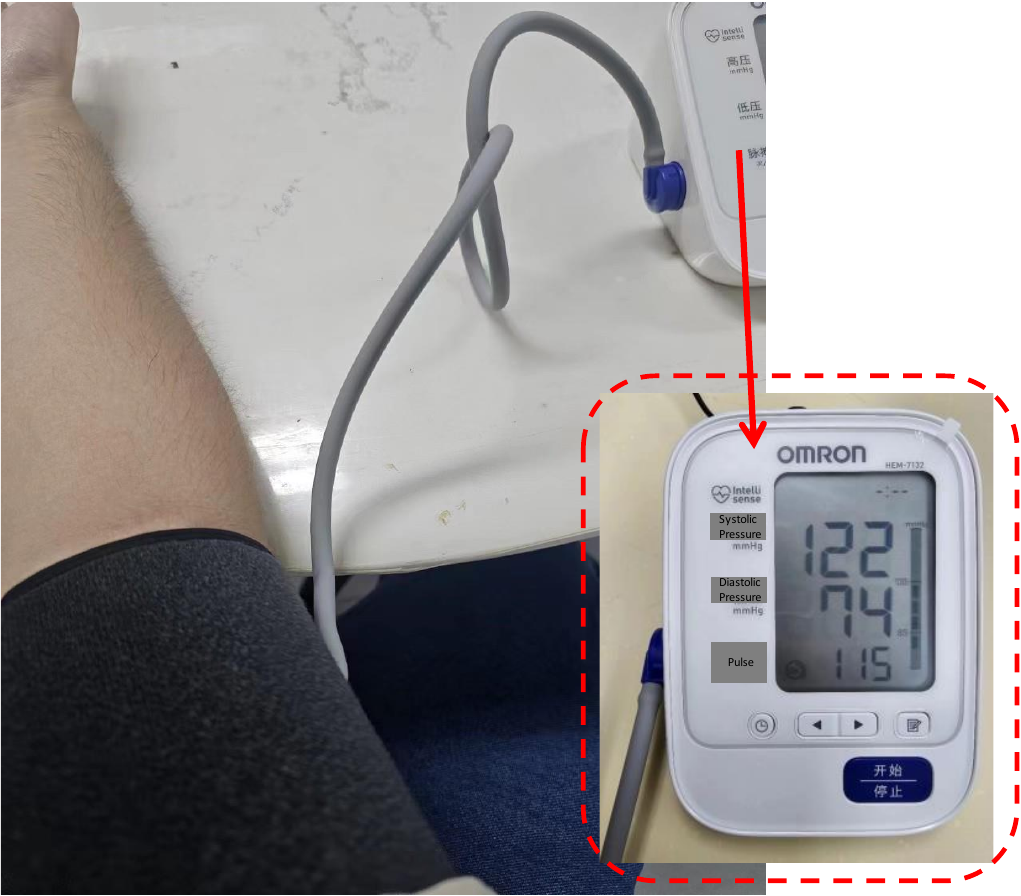}
        \caption{}
    \end{subfigure}
    \hfill
    \caption{A commercial BP monitor setup in a real clinic environment.}
    \label{commercial_Setup}
    \vspace{-0.1in}
\end{figure}

(2) Evaluation on representative commercial devices: 
Any changes in the timing rate of the BP monitor timing counter can cause measurement errors in systolic and diastolic blood pressure (see Appendix \ref{analysis_bp_monitor} for details). We conduct the experiment in a real clinic environment, in which an Omron HEM-7132 BP monitor is placed on a table in the waiting area outside the clinic diagnosis room.  The table is made of polyethylene with a thickness of 1.5cm. We position the transducer 10cm horizontally from the center of the backend enclosure of the monitor and place the magnetic field probe below the RTC crystal oscillator. The volunteer sits at the table like a regular patient, placing the arm on the table to measure blood pressure. The real clinic experiment scenario is shown in Fig. \ref{commercial_Setup}(a).

\begin{figure}[t]
\centering
\includegraphics[width=0.5\textwidth]{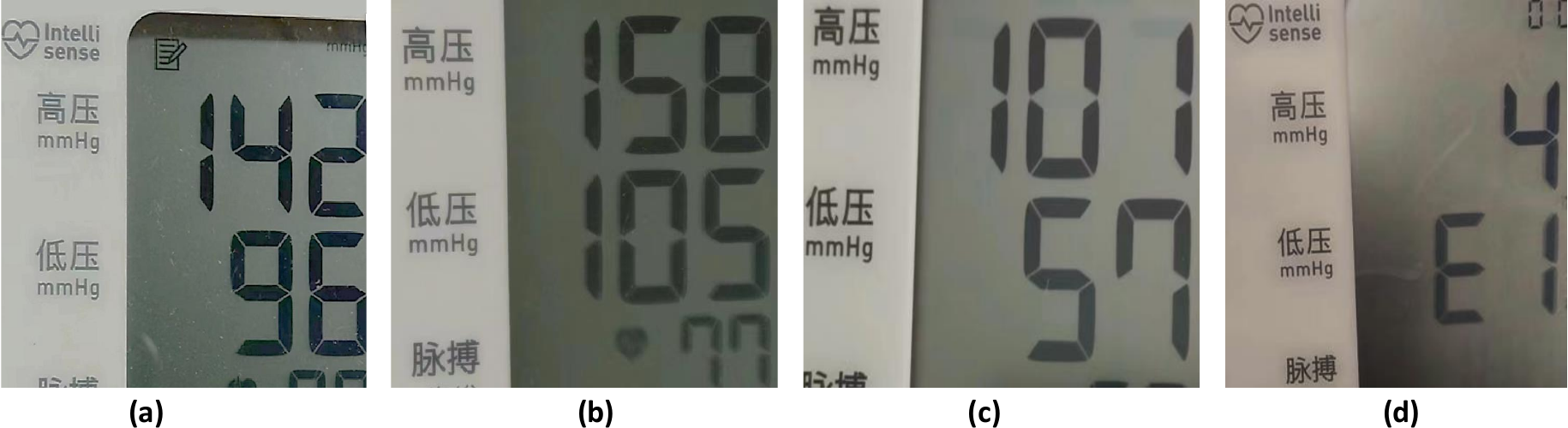}
\caption{Experiment results on HEM-7132.}
\label{exp_res1}
\vspace{-0.1in}
\end{figure}

Before launching the attack, the volunteer's blood pressure is measured at 122/74 mmHg (Fig. \ref{commercial_Setup}(b)). When the EM signal phase is observed to be $\pi/2$, we transmit an excitation signal of 98V, 2.16 rads with an interval of 0.014 ms, and repeat the test five times. We observe that the average measured blood pressure of the volunteer decreases to 101/57 mmHg (Fig. \ref{exp_res1}(c)), with systolic and diastolic pressures decreasing by 17\% and 23\%, respectively, and the cuff deflation speed slows down. When we emit continuous attack signals, the cuff nearly stops deflating, and the screen displays an E1 error (Fig. \ref{exp_res1}(d)). Meanwhile, we observe that the amplitude of the oscillating signal decreases from approximately 100 mVpp to 77 mVpp and remains stable, indicating that the phase of the induced electrical signal is $\pi$ out of the phase with the oscillating signal, causing the timing rate $f$ decreasing nearly to 0Hz and the valve then stops deflating (i.e, $v' \rightarrow 0$). Based on the established $\varphi \rightarrow \beta_1$ mapping, we fix the excitation signal voltage at 100V and adjust the signal phase to 2.3 and 2.31 rads, while maintaining a transmission interval of 0.01 ms. We report that the cuff deflation speed gradually increases, and the measured blood pressures are 142/96 mmHg (Fig. \ref{exp_res1}(a)) and 158/105 mmHg (Fig. \ref{exp_res1}(b)), respectively. For every 0.01 rads increase in phase, the systolic and diastolic pressures increase by approximately 11\% and 9\%, respectively.

We further follow the same testing procedures and attack device setup on a real Kaidianbao POS machine scenario. We observe that when an excitation signal with a phase difference of $\pi$ relative to the oscillating signal is transmitted, the timestamps recorded in the database for two consecutive card swipes within a few seconds are identical. With a constant transmission interval, a phase shift of 0.005 radians in the attack signal results in a forward time drift of approximately 5 seconds. Since both kinds of devices leverage timestamps to sense physical signals/perform operations, we envision that any device with those sensing/operating mechanisms will be vulnerable to TimeTravel, posing severe security risks.
% For illustrative commercial device examples, when we set the phase drift to 2.29 rads and emit 92V excitation signals, both the blood pressure and pulse values measured by the BP monitor increase significantly, as shown in Fig. \ref{commercial_devices}(a,b). When we apply an excitation signal with a phase drift of 2.27 rads and an amplitude of 89V to the POS machine and swipe the card every 6 seconds, the transaction information read from the database is shown in Fig. \ref{commercial_devices}(c). We observe that the time difference between the two recorded transactions is 16 seconds.

\begin{figure}[t]
\vspace{-0.1in}
    \centering
    \begin{subfigure}[t]{0.23\textwidth}
        \centering
        \includegraphics[width=\textwidth]{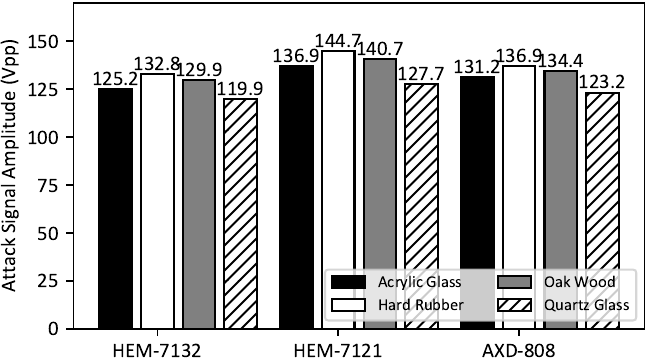}
        \caption{Blood Pressure to 180mmHg.}
    \end{subfigure}
    \hfill % Creates horizontal space between the images
    % \begin{subfigure}[t]{0.22\textwidth}
    %     \centering
    %     \includegraphics[width=\textwidth]{pics_and_tables/pic_7.pdf}
    %     \caption{}
    % \end{subfigure}
    % \hfill
    \begin{subfigure}[t]{0.23\textwidth}
        \centering
        \includegraphics[width=\textwidth]{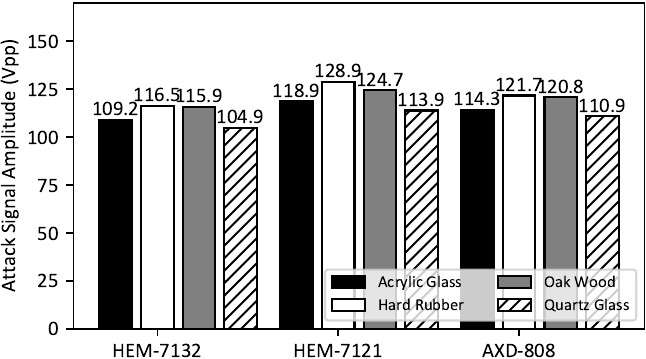}
        \caption{Blood Pressure to 90mmHg.}
    \end{subfigure}
    \hfill
    \caption{The minimum energy required for different BP monitors to attack under different solid materials.}
    \label{fig:impmaterials}
\end{figure}

\begin{figure}[t]
    \centering
    \begin{subfigure}[t]{0.2\textwidth}
        \centering
        \includegraphics[width=\textwidth]{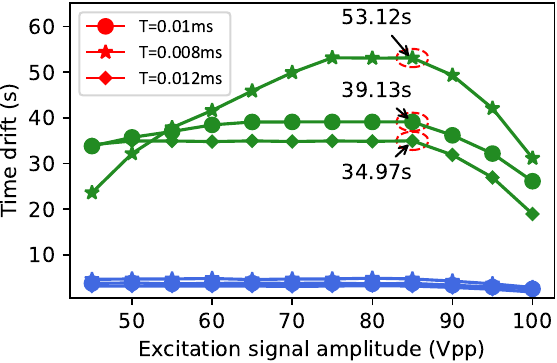}
        \caption{$\Delta=\frac{\pi}{12}$ (blue line) and $\frac{11\pi}{12}$(green line).}
    \end{subfigure}
    \hfill
    \begin{subfigure}[t]{0.2\textwidth}
        \centering
        \includegraphics[width=\textwidth]{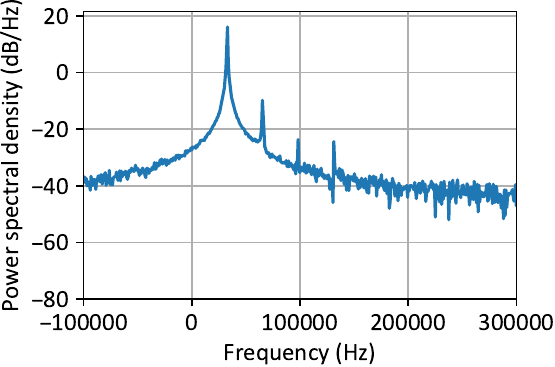}
        \caption{}
    \end{subfigure}
    \hfill
    \caption{Experimental results on forward drift of time.}
    \label{timeforwards_validate}
    \vspace{-0.2in}
\end{figure}

\textbf{B. Impact of Medium Materials.} The material of the solid medium plays a crucial role in the dispersion characteristics of waves and the amplitude of vibrations perpendicular to the medium's surface. To investigate the correlation between the minimum amplitude of the excitation signal and a successful attack, 
% we employ the DS1302 module,
we first employ three representative modules/development boards listed in Table \ref{moduletable} and place them respectively on four different mediums of \SI{5}{\mm}-thickness, including acrylic glass, hard rubber, oak wood, and quartz glass. We place the transducer \SI{20}{\cm} away from the RTC crystal, set the corresponding initial phase of the excitation signal for the backward drift of \SI{25}{\second} and the forward drift of \SI{45}{\second}, then quantify the minimum amplitude of the excitation signal necessary for successful attacks. For each experimental setting, we test and identify the attack signal phase that reduces the oscillation signal amplitude to its lowest level in advance, thereby establishing an $\varphi \rightarrow \beta_1$ mapping relationship. The results are listed in Table \ref{impact_materials}, showing that for different boards, the average energy required for successful implementation of time forward and backward drift attacks on the surface of the quartz glass medium is the smallest; the energy required for attacks on the surface of the hard rubber plastic is relatively the highest. Meanwhile, we adopt three representative BP monitors placed on the surface with the same attack environment configuration, and then record the minimum attack signal amplitude required to modify the high pressure to 180 and 90mmHg. The blood pressure of the tested volunteer in the resting state is 121/80mmHg. As the results in Fig. \ref{fig:impmaterials} show, the energy required to launch a successful attack does not differ much among different mediums in general,
%Results confirm 
 even though different mediums exhibit different velocities of transverse and longitudinal waves (see Table \ref{mediumdata}).
% indicating that as the velocities of transverse and longitudinal waves in the medium decrease, the energy required for mounting a successful attack increases in general.} 

\begin{table}[t]
\centering
\begin{tabular}{|c|c|c|c|c|}
\hline
\textbf{Material}&$\downarrow$\textbf{25s}&\textbf{Phase}&$\uparrow$\textbf{45s}&\textbf{Phase} \\ \hline
\multicolumn{5}{|c|}{\textbf{RTC Module: DS1302}} \\ \hline
Acrylic Glass& 75V&2.519& 68.2V&1.979 \\ \hline
Hard Rubber& 86V &2.018& 79.1V&1.471 \\ \hline
Oak Wood & 80V&1.972& 75.3V &1.426\\ \hline
Quartz Glass & 69V&2.299&  66.1V&1.753\\ \hline
\multicolumn{5}{|c|}{\textbf{Development Board: XC6C6SLX16}} \\ \hline
Acrylic Glass& 95V&2.518& 75.1V&2.395 \\ \hline
Hard Rubber& 104.2V & 2.132 & 85V&1.613 \\ \hline
Oak Wood & 99V& 2.213& 81.6V &1.728\\ \hline
Quartz Glass & 87.5V& 2.302&  71V& 1.897\\ \hline
\multicolumn{5}{|c|}{\textbf{Development Board: STM32-F407ZGT6}} \\ \hline
Acrylic Glass& 86V&2.536& 71.9V&2.413 \\ \hline
Hard Rubber& 94.9V &1.947& 81.2V&1.563 \\ \hline
Oak Wood & 90V&2.119& 77V &1.718\\ \hline
Quartz Glass & 79.8V&2.387& 68.5V & 2.012\\ \hline 
% \hline
% \textbf{Material}&\textbf{180mmHg}&\textbf{Phase}&\textbf{90mmHg}&\textbf{Phase} \\ \hline
% \multicolumn{5}{|c|}{\textbf{BP Monitor: Omron HEM-7132}} \\ \hline
% Acrylic Glass& 114.3V&1.482& 125.5V&1.697 \\ \hline
% Hard Rubber& 125.7V &1.559& 134.2V&1.782 \\ \hline
% Oak Wood & 117.9V&1.594& 129.3V &1.818\\ \hline
% Quartz Glass & 103.9V&1.612& 118.8V & 1.854\\ \hline 
% \multicolumn{5}{|c|}{\textbf{BP Monitor: Omron HEM-7121}} \\ \hline
% Acrylic Glass& 119.3V&1.649& 129V &1.828 \\ \hline
% Hard Rubber& 127.4V &1.563& 136.7V&1.747 \\ \hline
% Oak Wood & 120V&1.498& 130.8V &1.695\\ \hline
% Quartz Glass & 108.1V &1.712& 121.2V & 1.945\\ \hline 
\end{tabular}
\caption{The minimum amplitude of the excitation signal, as well as corresponding initial phases (deg) necessary for successful attacks, under different expected time drift.}
\label{impact_materials}
\vspace{-0.1in}
\end{table}

\begin{figure}[t]
    \centering
    \begin{subfigure}[t]{0.231\textwidth}
        \centering
        \includegraphics[width=\textwidth]{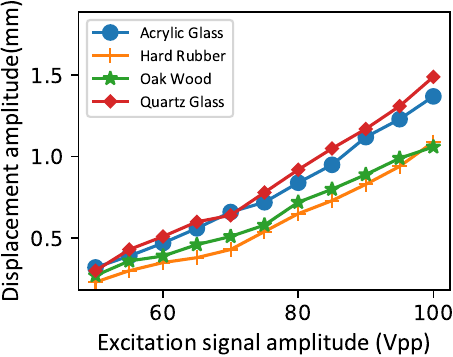}
        \caption{}
    \end{subfigure}
    % \hfill % Creates horizontal space between the images
    \begin{subfigure}[t]{0.236\textwidth}
        \centering
        \includegraphics[width=\textwidth]{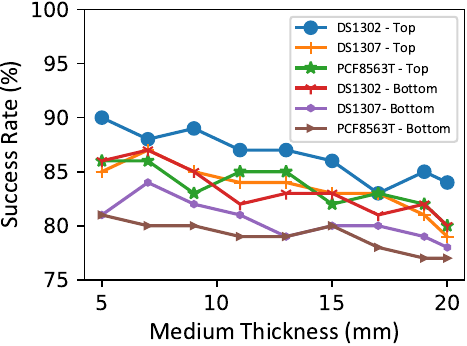}
        \caption{Acrylic Glass Surface.}
    \end{subfigure}
    \begin{subfigure}[t]{0.234\textwidth}
        \centering
        \includegraphics[width=\textwidth]{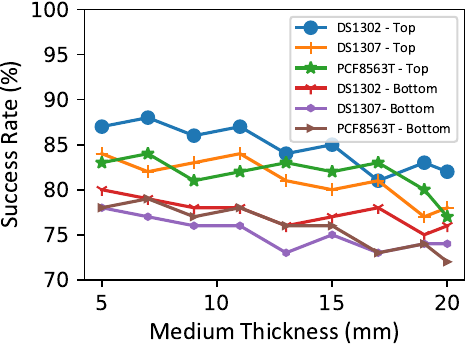}
        \caption{Hard Rubber Surface.}
    \end{subfigure}
    \begin{subfigure}[t]{0.234\textwidth}
        \centering
        \includegraphics[width=\textwidth]{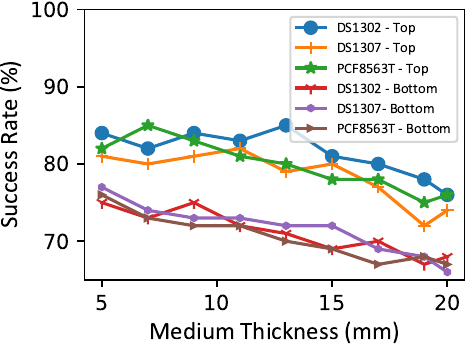}
        \caption{Oak Wood Surface.}
    \end{subfigure}
    \hfill
    \caption{The variation of surface vibration amplitude of the medium with the amplitude of the excitation signal (a); The success rate of attacks under different thicknesses of the medium and the positions of the transducer (b)-(d).}
    \label{impact_bc}
    \vspace{-0.2in}
\end{figure}

To further evaluate the relationship between the amplitude of the excitation signal and the surface vibration response of the medium, we apply excitation signals of different amplitudes to the transducer and place the GFL-Z30N-RS485 vibration meter close to the surface of the selected medium to measure the vibration amplitude at which the crystal located. The variation of the vibration amplitude on the surface of the medium is shown in Fig. \ref{impact_bc}(a). The results show that as the amplitude of the excitation signal increases, the mechanical vibration amplitude on the surface of the medium follows an overall upward trend. When the excitation voltage increases to 50V, the vibration amplitude of different mediums at the selected positions increases by more than 400\%. The difference in the vibration amplitude of different mediums under the same excitation signal may be caused by different structural characteristics (e.g., density and elasticity). For a medium with low density and high elasticity, attackers may need a larger signal amplitude to achieve successful attacks.

\textbf{C. Impact of Attack Energy on Time Drifting Forward.}
% According to Eq. \ref{eq:eqphase}, apart from the phase of superimposed electric signal, the variation in the intensity of the excitation signal can affect the amount of time drift. To validate the possible effect, we integrate DS1302 RTC module, a 5 mm-thick acrylic glass plate and place the transducer 3 cm next to the DS1302 RTC crystal, and set the amplitude of the signal driving the transducer to (45, 50, 55, 60, 65, 70, 75, 80, 85, 90, 95, 100Vpp), phase shift offset $\Delta$ to $\frac{\pi}{12}$ and $\frac{11\pi}{12}$. We start to emit signal when observe the oscillation signal emitted by the oscillator has a phase of $\frac{\pi}{2}$, then record the average amount of time-forward drift, as shown in Fig. \ref{timeforwards_res}(a). 
According to Eq. \ref{eq:eqphase}, apart from the phase of the superimposed electric signal, the variation in the intensity of the excitation signal can affect the amount of time drift. To validate this effect, we integrate a DS1302 RTC module with a 5 mm-thick acrylic glass plate, place the transducer 3 cm next to the DS1302 RTC crystal, and set the amplitude of the signal driving the transducer to various levels (45, 50, 55, 60, 65, 70, 75, 80, 85, 90, 95, 100Vpp). The phase shift offset $\Delta$ is set to $\frac{\pi}{12}$ and $\frac{11\pi}{12}$. We start emitting the signal when the oscillation signal from the oscillator has a phase of $\frac{\pi}{2}$ and then record the average amount of time-forward drift, as shown in Fig. \ref{timeforwards_res}(a). Under the setting of a single attack duration of \SI{0.008}{\ms}, when the attack signal energy increased from 65Vpp to 75Vpp, the time drift increased from \SI{45.89}{\second} to \SI{53.12}{\second}, approximately 15.8\%. At attack signal amplitudes of 75$\sim$85Vpp, we observe no significant change in time drift. Simultaneously, we use the probe to detect the oscillation signal and observe that the frequency band overlaps with the center frequency at the end of a single attack of different durations, proving that the oscillation signal can converge within the duration of a single attack. Therefore, increasing the amplitude to accelerate the phase drift speed cannot help obtain more drift time.

With the increase of the energy of the excitation signal, the forward drift time starts to decrease at 90Vpp. Fig. \ref{timeforwards_res}(b) shows the power spectral density of the amplitude normalized oscillation signal at the excitation signal amplitude of 100Vpp.
We observe a Lorentz-shaped spectrum broadening near the frequency of \SI{32.768}{\kilo\hertz}, with several harmonics based on  \SI{32.768}{\kilo\hertz} occurring in the power spectrum. We believe that it is due to the excessive excitation amplitude on the crystal oscillator, resulting in its nonlinear behavior: the harmonic out-of-band components in the circuit will reduce the amplification gain of the amplifier in the feedback circuit for the  \SI{32.768}{\kilo\hertz} in-band signal\cite{jsnfceraser},  making it difficult for the amplitude of the oscillator output signal to stably exceed the threshold of decreasing excitation counter values, which will lead the time not to drift forward as expected.

\textbf{D. Impact of Medium Thickness.}
The intensity of Lamb waves decreases rapidly with an increase in the thickness of the medium, affecting the efficiency of energy transfer. We use acrylic glass, hard rubber, and oak wood materials with thicknesses ranging from 5$\sim$20 mm. Placing the transducer on both sides of the board at the same position, we fix the distance between the transducer and the crystal oscillator at 20 cm. For each attack setting, we first establish $\varphi \rightarrow \beta_1$ through testing the phase response of oscillation signal, and find the attack signal phase that reduces the oscillation signal amplitude to its lowest level. Then, we adjust the phase of the excitation signal for a forward drift of 45 seconds and a backward drift of 25 seconds, maintaining the amplitude of the excitation voltage at 80Vpp and 85Vpp, respectively. The attack success rates with varying thicknesses of the plate are shown in Fig. \ref{impact_bc}(b), (c), and (d). As the thickness increases, the attack efficiency slightly decreases. When the thickness increases by 15 mm, the average success rate decreases by 7\% across different materials. Additionally, placing the transducer on the back of the board results in an average success rate decrease of about 6\% compared to placing it on the front. This is due to energy loss when sound waves transmit from the lower to the upper surface, reducing the mechanical energy that reaches the crystal oscillator.

\begin{figure}[t]
\vspace{-0.1in}
\centering
% \centerline{\includegraphics[width=0.3\textwidth]{pics_and_tables/pic_17.pdf}}
\begin{subfigure}[t]{0.22\textwidth}
        \centering
        \includegraphics[width=\textwidth]{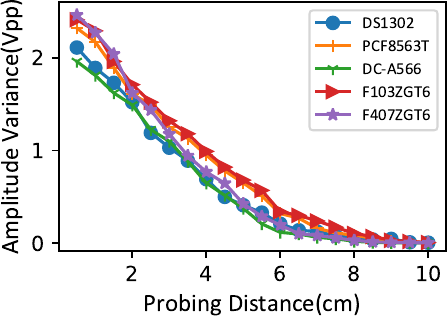}
        \caption{}
    \end{subfigure}
    \begin{subfigure}[t]{0.24\textwidth}
        \centering
        \includegraphics[width=\textwidth]{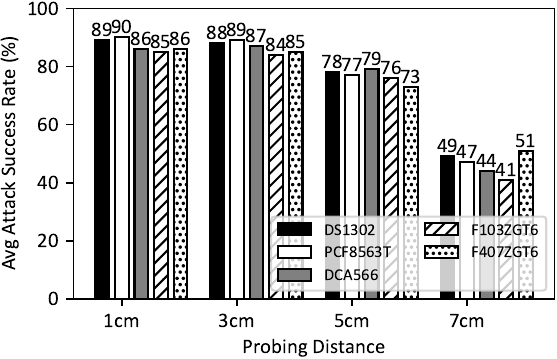}
        \caption{}
    \end{subfigure}
\caption{The variation in the peak-to-peak amplitude of the detected oscillation signal and the average attack success rate at different detection distances.}
\label{impact_delay}
\vspace{-0.2in}
\end{figure}

% \textbf{E. Impact of the Magnetic Sensing Distance.} The magnetic field signal's amplitude diminishes at a cubic rate with the distance increase from the source of radiation. Accurately tracking the phase of oscillation signals in the circuit is crucial for successfully mounting attacks. However, if the magnetic field probe is positioned too far from the crystal oscillator, the amplitude variance of the detected signal will not be well noticeable, making it difficult to track the phase of signals. We select three representative modules from Table \ref{moduletable}. We place the probe at different distances below the RTC crystal, and then probe the peak-to-peak amplitude variation, as shown in Fig. \ref{impact_delay}. As the probing distance increases, the amplitude variation of the signal gradually decreases, until at a distance of approximately \SI{7.5}{\cm}, the amplitude changes of the magnetic signal all decrease to below 0.08Vpp, and we can hardly recognize clear trends of amplitude changes under the presence of electromagnetic noise. However, in general
% %many potential attack 
% target devices are usually placed on solid surfaces with a thickness of no more than \SI{7.5}{\cm} (e.g., desktops, racks).
% %and we believe that 
% Therefore, under normal conditions, TimeTravel is capable of being effective in a large number of application scenarios.

\textbf{E. Impact of the Magnetic Sensing Distance.} We select five representative modules listed in Table \ref{moduletable} and place them respectively on a 5 mm-thickness acrylic glass plate. We then place the transducer 20 cm away from the RTC crystal, set the corresponding initial phase of the excitation signal for the backward drift of 25 s and the forward drift of 45 s. We position the magnetic probe at various distances below the RTC crystal oscillator and automatically track the moments when TimeTravel recognizes the phase of an oscillation signal as $\pi/2$, at which point the corresponding excitation signal is emitted. For each module, we repeat the attack for 100 times, and the observed peak amplitude of the oscillation signal and the average attack success rates are shown in Fig. \ref{impact_delay}(a) and (b), respectively. As the probing distance increases, the amplitude variation of the signal gradually decreases, until at a distance of approximately \SI{7.5}{\cm}, the amplitude changes of the magnetic signal all decrease to below 0.2Vpp, and we can hardly recognize a clear trend of amplitude changes under the presence of electromagnetic noise. Meanwhile, when the probe distance from the crystal oscillator is greater than 5cm, we observe that the attack success rate decreases significantly. At a distance of 7cm, the average success rate is only 46\%. Due to the small amplitude change of the detected oscillating signal, when the attack distance is 7cm, TimeTravel wrongly misjudges other phases as $\pi/2$ for many times, resulting in the wrong time drift amount.

\subsubsection{EM Side-channel Quality}
To assess the quality of the electromagnetic side-channel, we use commercial BP monitors described in Section 6.1.2 with enclosures, as well as all modules and development boards listed in Table \ref{moduletable}. We place each device under test on a 5mm thick acrylic glass surface, positioning the magnetic field probe 1$\sim$4 cm below the MCU on the device's PCB board. We collect electromagnetic signal sequences of 100ms per data sample for the training set. For the testing set, we place the probe at distances of 1$\sim$7 cm directly below the MCU of another batch of devices, collect 100ms electromagnetic signal sequences, and conduct classification tasks. For the parameter settings, we set $a=b=50$ and $M=5$MHz. The accuracy, recall rate, and F1 score of the model for the six BP monitors are shown in Table \ref{modeldetectres}. Generally, for time series samples with a detection distance of 1$\sim$4 cm, TimeTravel achieves a classification accuracy of more than 92\% across different solid material surfaces. However, as the detection distance increases, the classification accuracy significantly decreases accordingly. At a 7 cm detection distance, the average classification accuracy for the three solid materials drops to below 40\%. We observe that this decline corresponds with the average SNR of the test data decreasing from 21dB at a 1 cm detection distance to 6dB at a 7 cm detection distance, making it harder for the neural network to accurately capture spectral features and correctly classify the device model.

\subsubsection{Processing Delay}
TimeTravel captures and analyzes the signals emanating from the oscillator circuit in order to launch the attack at the appropriate times, which introduces additional processing delay, spanning from the moment the target phase signal is captured from the oscillator circuit to the beginning of the emission of the acoustic attack signal. We use modules listed in Table \ref{moduletable} and a laptop equipped with a 13th generation Core i7-1355U processor and 32 GB of RAM, running on the Debian 12 operating system, to measure the processing delay. We keep the attack settings consistent with Section \ref{section_succrate} (A) and then repeat the attack 100 times with the time drift backward by \SI{5}{\second} and \SI{25}{\second}, as well as forward by \SI{25}{\second} and \SI{45}{\second}. We record the hardware timestamps at which USRP begins to perceive the target phase of the oscillation signal and begins to transmit the signal. We then calculate the average delay, which is listed in Table \ref{tabledelay}. The average delay for different modules ranges from \SI{0.77}{\micro\second} to \SI{0.87}{\micro\second}, meaning that under normal circumstances, attackers can mount attacks with a processing delay of no more than approximately 2.9\% of oscillation cycles (the oscillation cycle is around \SI{30.52}{\micro\second}).

\subsubsection{Robustness}
In real-world scenarios, a target device may be placed with extraneous items nearby. To evaluate the attack robustness of TimeTravel in the presence of extraneous items, 
we use a polyethylene plastic table with a thickness of \SI{1.5}{\cm} (covered with spray paint), 
place the transducer about \SI{50}{\cm} away from the RTC crystal, and adjust the amplitude and phase based on the excitation signal. We set up two scenarios, and in each scenario, we have different item layouts around the module, as shown in Fig. \ref{ac_scenarios_setting}. 

For setting 1, we place a digital camera, an electronic fan, a cup, a box, wood bricks, and a gateway on the table. For setting 2, we place a lamp, an electronic fan, a cup, and wood bricks on the table. Moreover, we set up a baseline without placing any extraneous items on the table to compare changes in attack efficiency.  During the test, we keep the fan, digital camera, gateway, and lamp running, to maintain the presence of noise from other devices.
% For settings 1 and 2, we keep the fan, digital camera, and lamp running, to simulate the presence of noise from other devices.
% For setting 3, we place additional four A4-sized sheets of paper.
%Note that Fig. \ref{ac_scenarios_setting} presents the two scenarios under the test of the STM32F407ZGT6 module, and we only replace the module and keep the placement of the other items the same for the tests with different modules. 
We set the transducer \SI{30}{\cm} away from the RTC crystal, the amplitude of the excitation signal to 86V, and repeat the attack 100 times with forward drift by \SI{45}{\second} and then measure the average success rate under each representative module, which is shown in Table \ref{res_robustness}.
We observe that when extraneous items are placed around the module, there is no significant change in the success rate of the attack compared to the baseline. The vibration waves of different frequencies propagate independently in the medium and do not interfere with each other, and only the acoustic vibration waves emitted by attackers will affect RTC timing. However, since TimeTravel relies on the conduction of mechanical energy between solid mediums, placing extraneous items beneath the target module may reduce the efficiency of the energy transfer, resulting in a lower attack success rate. The thicker the items, the lower the energy transfer.
% The thicker the object, the lower the energy transfer efficiency and therefore the lower the attack efficiency.
% We observe that when there are other items below the module, the success rate of the attack is reduced by about 5\% compared to when no unrelated items are placed below; When only extraneous items are placed around the module, we observe no significant change in the success rate of the attack compared to the attack success rate without any irrelevant items around it (we refer it as Setting 4). We believe that extraneous items beneath the module will reduce the efficiency of the energy transfer from the surface of the medium to the crystal, resulting in a lower attack success rate. Vibration waves of different frequencies propagate independently in the medium and do not interfere with each other, only the acoustic vibration waves emitted by attackers will affect RTC timing.
%\vspace{-0.1in}
\section{Discussion}
In this section, we discuss the potential countermeasures, safety recommendations, and impact of NTP synchronization.

\vspace{-0.1in}
\subsection{Countermeasures}
The key to defend against TimeTravel attacks lies in preventing acoustic vibrations from generating or propagating disruptive electrical signals in an RTC. Different from the potential 
hardware and software-based methods proposed in \cite{bolton2018blue},  we propose two kinds of countermeasures and have disclosed them along with the security threats to BP monitor vendors, including Omron, HYNAUT, Yuwell, and Xiaomi:

\textbf{A. Replacing Oscillating Source.} Apart from crystal, manufacturers can use frequency synthesizers\cite{manassewitsch1987frequency} or MEMS oscillators to generate $f_{output}$=32.768 kHz oscillation signals. For instance, Silicon Labs Si5351 clock generator\cite{clkgenerator} can generate specific frequency clock signals by using a 25MHz crystal oscillator that provides reference signals. Once powered with a 2.5V or 3.3V supply, the PLL and Multisynth divider parameters can be configured via the $\text{I}_2\text{C}$ interface according to Eq. \ref{ossourcefulmula}:
\begin{align}
    f_{PLL}&=\Delta_{PLL}\times 25\text{MHz} \nonumber \\
    f_{output}&=\frac{f_{PLL}}{\Delta_{multisynth}},
    \label{ossourcefulmula}
\end{align}
where $\Delta_{PLL}$ and $\Delta_{multisynth}$ can be determined as 36 and 27465.82 (fractional division), respectively, resulting in $f_{output}$ as 32.768kHz. Finally, Si5351 outputs the generated clock signal through the CLK0 pin to the clock input pin of the RTC MCU. 
Generating 25MHz ultrasonic waves is challenging because of their weak diffraction ability. These waves typically propagate only a few millimeters in most solids, making it highly unlikely for a 25MHz crystal oscillator to be disturbed by acoustic signals.
% Generating ultrasound at 25MHz is typically challenging and requires specialized transducers and substantial power. In solid mediums, even if high-frequency ultrasound is successfully generated, it attenuates rapidly during propagation. The rate of attenuation, known as the attenuation coefficient $\beta$, is generally proportional to the frequency $f:\beta=\beta_0 f$.}
% \textcolor{red}{ $\beta_0$ is determined by the intrinsic properties of the solid materials (see detailed values in \cite{ono2020comprehensive}), and $f$ refers to acoustic frequency. Also, it indicates that the maximum propagation distance is independent of the initial amplitude of the signal. Considering the relationship between the attenuation coefficient and frequency in solid materials, the maximum propagation distance can be estimated by (i.e., the distance at which the signal amplitude decreases to $1/e$ of its original value):}
% \begin{align}
%     \text{Maximum Propagation Distance}=\frac{1}{\beta_0 f}.
% \end{align}

% \textcolor{red}{Given materials listed in Table \ref{mediumdata}, the maximum propagation distances are approximately between 0.033$\sim$0.2 cm. Considering that the typical thickness of a solid surface holding devices (e.g., desktop) is usually greater than this range, it is challenging for the signal to propagate from the underside to the topside of the solid surface. This means that even if the transducer is placed directly underneath the RTC module, it is unlikely to interfere with the 25MHz crystal by transmitting the signal through the solid material.}

\textbf{B. Applying Shock-absorbing Materials.}
To reduce interference from potential resonance signals on the oscillator's vibration mode, manufacturers can use shock-absorbing materials to cover the timing circuit PCB containing the \SI{32.768}{\kilo\hertz} crystal oscillator. For RTC circuits and mechanical vibrations perpendicular to the plane, this scenario can be modeled as a mass-spring-damper system. The energy transfer function $H(\omega)$ of this system is given by\cite{rao1995mechanical}:
\begin{align}
H(\omega)=\frac{1}{\sqrt{(1-(\frac{\omega}{\omega_n})^2)^2+(2\zeta\frac{\omega}{\omega_n})^2}},
\end{align}
where $\omega$ is the frequency of the excitation signal, $\omega_n$ is the natural frequency of the material, $\zeta=\frac{c}{2\sqrt{km}}$ is the damping ratio, $c$ and $k$ are the damping coefficient and stiffness of the damping material, respectively, and $m$ is the equivalent mass of the device and solid surface. As the damping coefficient $c$ increases, 
% the damping ratio $\zeta$ also increases, causing $H(\omega)$ to exhibit a decrease in the peak value of the vibration signal response. This makes the damping material's response more even across a range of frequencies. 
the damping material's response becomes more even across a range of frequencies.
Therefore, a higher damping coefficient results in less energy reaching the RTC crystal oscillator, providing more effective protection against mechanical vibration interference. Additionally, since mechanical vibrations are transmitted perpendicularly to the surface, placing the damping material perpendicular to the direction of vibration allows it to directly dissipate vibrational energy. However, when the damping material is placed at an angle, the vibrational energy is partially converted into shear force and partially into compressive force, reducing the effectiveness of energy absorption. Manufacturers can use damping materials with relatively high damping coefficients, such as Polyurethane Foam and Polymeric Viscoelastic Materials, and place the damping material parallel to the solid plane to maximize the absorption of mechanical vibrational energy.

\vspace{-0.1in}
\subsection{Safety Recommendations}
When launching the attack, the electrical signal amplitude applied to the transducer often exceeds the maximum safe voltage for human exposure (36V). Therefore, it is strongly recommended that researchers wear insulating gloves during testing to prevent electric shock and use insulating tape to secure the transducer and its connected wires.

Lamb waves propagating in solids do not directly impact the human body as they are primarily confined within the material. However, these waves can interact with the air at the material boundaries, generating sound waves with the same or similar frequencies. Fortunately, such a process shows low energy conversion efficiency, due to the significant acoustic impedance difference at the solid-liquid interface. In our experiments, the detected sound pressure level at the material surface is no higher than 26 dB, comparable to ambient noise levels. However, to avoid potential health risks, researchers are advised to control the intensity of Lamb waves, ensuring that they remain within safe sound pressure levels for human exposure, especially when using powerful electrical signals to drive the transducers. 
For the testing experiment of the blood pressure monitor, we have obtained approval from the Institutional Review Board (IRB) before the test, and the participant is aware of the entire process of the experiment and is informed of the potential impact upon human subjects.

\vspace{-0.1in}
\subsection{Impact of NTP Synchronization}
The commercial POS machine we test uses NTP only to get the current time at startup, updates the RTC timing, and then works only with the RTC internal timestamp. Therefore, the POS machine does not rely on NTP after a successful startup. Moreover, there are some devices whose RTCs are not synchronized via NTP, e.g., Industrial PLCs, smart electricity/gas/water meters, and medical devices.
NTP synchronizes system time periodically, often at intervals of minutes or hours based on device accuracy needs\cite{mills2003brief}. Altering RTC timing within a short span does not prompt NTP synchronization. However, in many scenarios, especially in real-time computing and communication systems, even temporary timestamp modifications can cause severe damages like service interruptions or cascade failures. Even if NTP syncs after or during the attack, damages are already done and irreversible.

\section{Related Work}
We survey previous research that utilizes acoustic signals to interfere with the operating logic of devices, which can be divided into two categories:
% The recent works related to acoustic signal attacks can be mainly divided into two categories: 
(1) acoustic command attacks using sensors' nonlinear interpretation of acoustic signals and (2) acoustic signal resonance jamming attacks. 

% For the first type of attack, prior work focuses on speech recognition algorithms for voice control systems that exploit the nonlinear response of microphone hardware to ultrasound to inject commands inaudible to a human ear into the target voice control system.
Prior work on the first type of attack targets voice control systems by using speech recognition algorithms to exploit microphone hardware's nonlinear response to ultrasound, injecting inaudible commands. Zhang et al. \cite{zhang2017dolphinattack} first exploited a hardware loophole in the nonlinear response of a microphone, to convert ultrasound waves into valid voice commands and manipulate several brands of voice assistants. Roy et al. \cite{roy2018inaudible} utilized speaker arrays to effectively increase the acoustic stealth attack distance by sending narrow-band ultrasound waves. Yan et al.\cite{yan2020surfingattack} proposed to inject ultrasonic commands into voice assistants through ultrasound waves propagating in a solid medium to 
%avoid the limitation of 
bypass physical spatial obstacles. Ji et al.\cite{ji2021capspeaker} utilized the feature that a capacitor emits a high-frequency noise when it is being charged or discharged, and configured specific voltages to the two sides of the capacitor to make the noise propagate to a cell phone's microphone to control the voice assistant to execute commands. Yang et al. \cite{yang2023remote} controlled the  voice assistant by manipulating MEMS switching power supply to make noise.

For the second type of attack, attackers exploit the resonant frequency of electronic components to affect the normal operation of specific equipment or modules. Son et al. \cite{son2015rocking} emitted acoustic signals that are close to the resonance frequency of the gyroscope inside a drone, causing the gyroscope to resonate and then making the drone's flight trajectory change. Trippel et al. \cite{trippel2017walnut} leveraged this feature to attack the output value of MEMS acceleration sensors. Bolton et al. \cite{bolton2018blue} emit acoustic resonance signals to the the magnetic head of a mechanical hard disk to make it positioned incorrectly, leading to the denial of service attack.
% found that by emitting acoustic signals with the same resonance frequency as the magnetic head of a mechanical hard disk, the magnetic head can be positioned incorrectly, leading to the denial of service attack.
Other methods are also proposed to perform acoustic signal injection attacks \cite{tu2018injected,dean2007degradation,carlini2016hidden,song2017poster}. In this study, TimeTravel first utilizes a ceramic transducer to emit sound vibration signals to the RTC crystal oscillator, causing the shift of the signal's phase and amplitude. 
Then, TimeTravel is able to modify the system time stored in the RTC counter, 
posing a serious threat to the reliability of devices that utilize system time for real-time execution.

%\vspace{-0.1in}
\section{Conclusion}
This paper presents a new security threat posed by a real-time RTC clock timing rate drift attack called TimeTravel. As a technique target for various real-time computing and communicating devices, TimeTravel leverages acoustic interference to modify the timing rate of the device's internal RTC clock, leading to the system timestamps drift.  
%by utilizing acoustic signal interference. 
By analyzing the response characteristics of quartz crystals to acoustic resonance signals of different amplitudes and phases, we reveal a quantitative relationship between the parameter settings of the acoustic resonance signals and the amount of RTC time drift. We validate the efficacy of TimeTravel on nine off-the-shelf modules with RTC circuits and seven commercial devices, and evaluate the robustness under two realistic placement settings. Finally, we propose countermeasures against TimeTravel.

% %-------------------------------------------------------------------------------
% \section*{Acknowledgments}
% %-------------------------------------------------------------------------------

% The USENIX latex style is old and very tired, which is why
% there's no \textbackslash{}acks command for you to use when
% acknowledging. Sorry.

%-------------------------------------------------------------------------------
\bibliographystyle{unsrt}
\bibliography{ref}

\appendix
\section*{Appendices}

\section{Verify the Feasibility of Transmitting Acoustic Waves in the Air to Interfere with Crystal Oscillator}
\label{validsoundpropfalse}

\begin{figure}[t]
    \centering
    \begin{subfigure}[t]{0.19\textwidth}
        \centering
        \includegraphics[width=\textwidth]{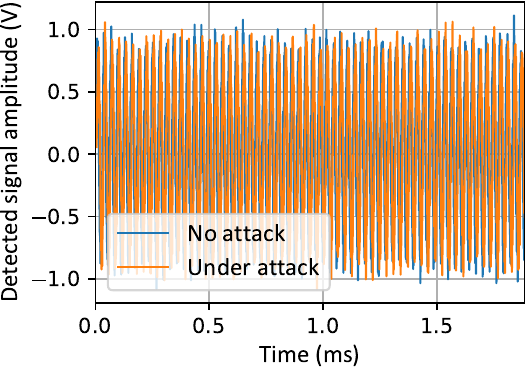}
        \caption{}
    \end{subfigure}
    \hfill % Creates horizontal space between the images
    \begin{subfigure}[t]{0.23\textwidth}
        \centering
        \includegraphics[width=\textwidth]{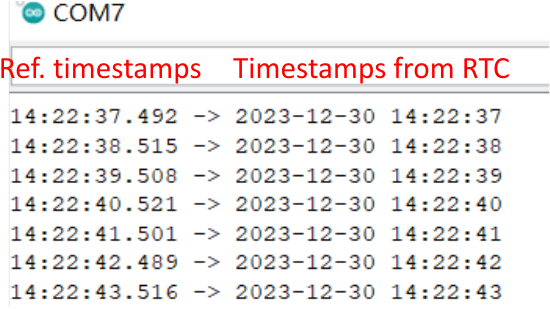}
        \caption{}
    \end{subfigure}
    \hfill
    \caption{Time domain of oscillation signal detected after sending 32.768kHz ultrasonic wave through the air(left); Serial output from development board (right).}
    \vspace{-0.2in}
    \label{soungpropair111}
\end{figure}

To verify the possibility of transmitting ultrasonic waves through the air to change the speed of RTC timing, we use the DS1302 RTC module and connect it to the Arduino UNO Rev3 development board through Dupont wire. Moreover, we use a pair of XHXDZ-5140 and XHXDZ-4140 ultrasonic speakers respectively, and place them approximately \SI{1.5}{\cm} on both sides of the RTC module. The optimal operating frequency band of XHXDZ-5140 is 10$\sim$\SI{30}{\kilo\hertz}, and the optimal operating frequency band of XHXDZ-4140 speaker is 26$\sim$\SI{46}{\kilo\hertz}. Through these two speakers, we are able to verify the sensitivity of the RTC module to interference from 20$\sim$\SI{46}{\kilo\hertz} ultrasonic acoustic signals. We use alligator clips to connect the ultrasonic speakers to the signal source and excite 20Vpp, 20$\sim$\SI{46}{\kilo\hertz} sinusoidal signals. We observe that the RTC time is still increasing forward at 1-second intervals. The oscillation signal emitted by the quartz crystal oscillator before and after the attack detected by the magnetic field probe is shown in Fig. \ref{soungpropair111} (a), and it can be seen that there is no significant change in the amplitude and frequency of the oscillation signal. We then adjust the transmission voltage and the distance between the speaker and the RTC module,  varying between 20$\sim$50Vpp and 0.5$\sim$\SI{1.5}{\cm}, respectively. We observe that the oscillation signal during the attack is still similar to that shown in Fig. \ref{soungpropair111}(a), and the RTC module maintains normal timing (Fig. \ref{soungpropair111}(b)). The failure of injecting attack signals over the air is due to the significant acoustic impedance and medium density disparity between fluids and solids. This inefficiency is compounded by the fact that ultrasound waves at the resonance frequency of 32.768 kHz have short wavelengths, making it difficult for them to penetrate the metal casing that encases quartz chips.

\begin{figure}[t]
\centering
\centerline{\includegraphics[width=0.28\textwidth]{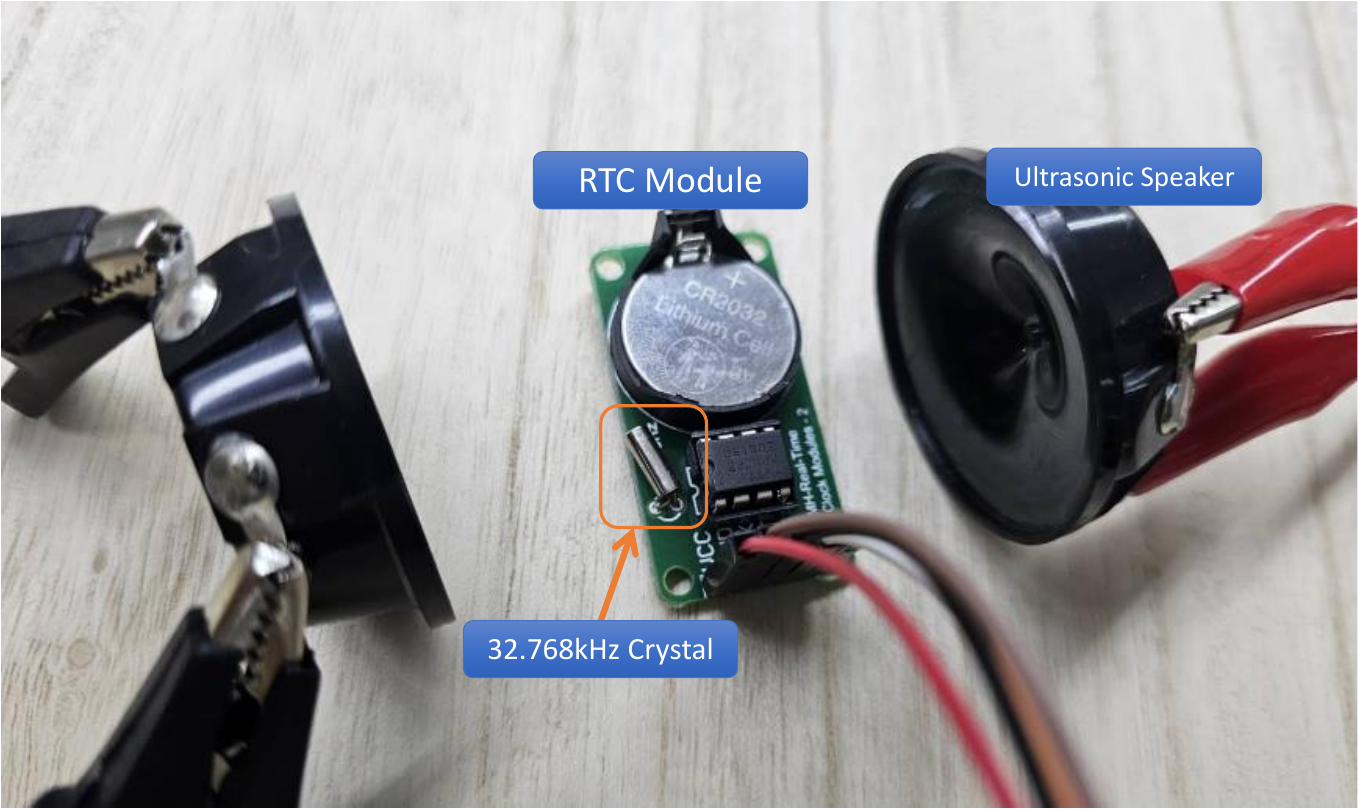}}
\caption{Experimental setup.}
\label{soungpropair}
\vspace{-0.1in}
\end{figure}

\section{The Propagation Characteristics of Lamb Waves}
\label{drlambwavefunc}
Lamb waves are elastic waves that propagate on the surface of a solid flat plate, consisting mainly first-order symmetric Lamb waves $S_x$ and first-order anti-symmetric Lamb waves $A_x$. The speed of Lamb wave propagation mainly depends on the frequency of the propagating wave and thickness of the solid medium\cite{mayer2011guided}. 

Suppose the excitation signals from transducer $f(t)=p\sin(\omega t+\varphi)$  successfully excites a Lamb wave with a frequency of $\omega$, when the first-order anti-symmetric wave component $u$ propagates along the $Z$ direction in a solid plate with the thickness of $2h$ (see Fig. \ref{propagation_dic}), under the excitation signal from sound source $V(t)$, the displacement equations of the particle at a certain position $(x,z)$ near the surface of the plate in the $X$ and $Z$ directions can be expressed as \cite{graff2012wave}:
\begin{align}
    % u_x(x,z,t)&=[ikA\cos(\alpha x)+\beta B \cos(\beta x))] \cdot V(t)\cdot e^{i(kz-\omega t)} \notag \\
    % u_z(x,z,t)&=[-\alpha A \sin(\alpha x)-ikB \sin(\beta x)]\cdot V(t)\cdot e^{i(kz-\omega t)} \notag \\
    u_x(t)&=[\alpha A \cos(\alpha x)-ik_aB \cos(\beta x)]\cdot V\cdot e^{i(\omega (t-t_T)+\varphi-k_az)} \notag \\
    u_z(t)&=[ik_aA\sin(\alpha x)-\beta B \sin(\beta x))]\cdot V\cdot e^{i(\omega (t-t_T)+\varphi-k_az)}, 
    \label{dispeq_xz}
\end{align}
where $V=p$ denotes the variation of amplitude of the excitation signals, $t_T=\frac{d}{c_s}$ is the time required for the wave to propagate to $(x,z)$, $t>t_T$; $d$ is the distance between the sound source and the location $(x,z)$; $c_s$ is the phase velocity, which can be calculated by Eq. \ref{eqset1}. A and B can be solved by making the coefficient determinant of the following equation set 0:
% \begin{equation}
%    \begin{bmatrix}
%   -2ik\alpha \sin(\alpha h)&(k^2-\beta^2)\sin (\beta h) \\
%   (k^2-\beta ^2)\cos (\beta h)&-2ik\beta \cos (\beta h) \\
% \end{bmatrix}
% \begin{pmatrix}
%  A\\
%  B
% \end{pmatrix}=
% \begin{pmatrix}
%  0\\
%  0
% \end{pmatrix} 
% \label{matrixeq1}
% \end{equation}
\begin{equation}
    \begin{bmatrix}
  2ik_a\alpha \cos (\alpha h)&(k_a^2-\beta ^2)\cos (\beta h) \\
  (k_a^2-\beta ^2)\sin (\beta h)& 2ik_a\beta \sin(\beta h)
\end{bmatrix}
\begin{pmatrix}
 A\\
 B
\end{pmatrix}=
\begin{pmatrix}
 0\\
 0
\end{pmatrix},
\label{matrixeq2}
\end{equation}
$\omega$ is the angular frequency of waves, $t$ represents the time, $k_a=\frac{\omega}{c_s}$ is the wave number of first-order anti-symmetric Lamb wave, $\alpha = \sqrt{(\frac{\omega^2}{c_L})^2-k_a^2}$,$\beta = \sqrt{(\frac{\omega^2}{c_T})^2-k_a^2}$, $c_L$ and $c_T$ represent the longitudinal and transverse wave speed, the corresponding values can be found in Table \ref{mediumdata}. 

For the displacement equation of anti-symmetric Lamb wave along $X$ direction, by using the prosthaphaeresis formula and Eq. \ref{superimpose_eq}, we can get the real part of 
 the $u_x$, which represents the observable physical displacement:
\begin{align}
    \xi = \mathrm{Re}\{u_x(t)\}= \lambda \sin(\omega t +\Phi),
\label{displacement_function}
\end{align}
where
\begin{align}
    \lambda & = \sqrt{[\alpha Ap\cos(\alpha x)]^2+[k_aBp\cos(\beta x)]^2} \notag \\
    \Phi & = -\omega t_T+\varphi-k_az+\arctan (\frac{\alpha A\cos(\alpha x)}{k_aB\cos(\beta x)}).
\end{align}

% \begin{align}
% \gamma & = \frac{p}{2}\{C\alpha\cos(\alpha x)\sin(\varphi+k_az-\omega t_T) \\
% & -k_aD\cos(\beta x)\cos(\varphi+k_az-\omega t_T)\} 
% \notag \\
%     \lambda & = \sqrt{[\frac{\alpha}{2}Cp\cos(\alpha x)]^2+[\frac{k_a}{2}Dp\cos(\beta x)]^2} \notag \\
%     \Phi & = -\omega t_T+\varphi-k_az+\arctan (\frac{\alpha C\cos(\alpha x)}{k_aD\cos(\beta x)}) \notag
% \end{align}

In practice, the propagation of sound waves in solids results in energy loss, which is reflected in a decrease in the amplitude of the waves. Therefore, $V$ should be multiplied by an energy loss ratio $\varepsilon_{\Delta z}$: $V'=\varepsilon_{\Delta z} V$, which can be determined by experimental testing.  For a certain solid medium, when $z$ is determined, $t_T=\frac{z}{c_s}$ is also determined, so attackers only need to determine $z$ and $\varphi$ to determine the phase of the displacement equation at a certain time. Since we focus on the particle displacement on the surface of the medium, so $x$ can be set to 0. Additionally, When Lamb waves are not applied and the plate, there may exist prestress in the surface of a plate, representing the deformation of the plate in steady state, so a constant $\gamma$ may be applied to $\xi$ (i.e., $\xi'=\gamma+\xi$).

\section{Derivation of Sinusoidal Signal Superposition Expression}
\label{dersupexpression}
Suppose both signals with same frequency are $f(t)=A\sin(2\pi ft+\beta_1)$ and $g(t)=B\sin(2\pi ft+\beta_2)$, respectively. According to the principle of vector superposition and sum-to-product formula, the result of the superposition of signals can be expressed as:
% \begin{eqnarray}\label{fgtemp1}
% f(t)+g(t)=&A\sin(2\pi ft+\beta_1)+B\sin(2\pi ft+\beta_2)  \nonumber \\
% =&A\sin(2\pi ft+\frac{\beta_1+\beta_2}{2}+\frac{\beta_1-\beta_2}{2})) \nonumber\\
% &+ B\sin(2\pi ft+\frac{\beta_1+\beta_2}{2}-\frac{\beta_1-\beta_2}{2})).
% \end{eqnarray}

% By using sum-to-product formula, Eq. \ref{fgtemp1} can be simplified as:
\begin{eqnarray}
  f(t)+g(t)=&(A+B)\cos(\frac{\beta_1-\beta_2}{2})\sin(2\pi ft+\frac{\beta_1+\beta_2}{2})\nonumber\\
&+(A-B)\sin(\frac{\beta_1-\beta_2}{2})\cos(2\pi ft+\frac{\beta_1+\beta_2}{2})\nonumber\\
=&\phi\sin(2\pi ft+\frac{\beta_1+\beta_2}{2}+\gamma),
\label{superimpose_eq}
\end{eqnarray}
where $\phi=\sqrt{A^2+B^2+2AB\cos(\beta_1-\beta_2)}$, and $\tan \gamma=\frac{A-B}{A+B} \tan(\frac{\beta_1-\beta_2}{2})$.
% \begin{align*}
%   \phi&=\sqrt{A^2+B^2+2AB\cos(\beta_1-\beta_2)} \\
%   \tan \gamma&=\frac{A-B}{A+B} \tan(\frac{\beta_1-\beta_2}{2}).
% \end{align*}

\section{Derivation of the relationship between mechanical stress and additional electrical resonance signals generated by crystal oscillators}
\label{derimechstress}
After being subjected to mechanical pressure, quartz crystal will generate inertial forces, causing it to vibrate within a linear range. Because one end of the crystal oscillator is fixed by electrodes and the other end vibrates freely, the mechanical vibration of the crystal oscillator can be summarized as a cantilever beam vibration problem under dynamic loads. Assuming the equivalent mass of the free end of the crystal oscillator is $M$; $S$ and $K$ are the equivalent damping constants and equivalent stiffness; $b,d$ are the width and the thickness of the quartz plate; $a$ is the acceleration caused by mechanical vibration at the free end of the crystal;$y$ is the displacement of the free end. The vibration process can be treated as a cantilever beam model\cite{zhang2004modelling}. Assuming that the overall acceleration caused by external mechanical excitation is $a(t)= A\sin(\omega t+\varphi)$, according to Newton's second law, the total force on the free end of the crystal flake(can be regarded as a cantilever beam) is the result of the combined action of driving force (generated by external vibration), restoring force and damping force:
\begin{align}
    My^{''}+Sy^{'}+Ky=MA\sin(\omega t+\varphi). \label{eq:diffeq}
\end{align}

By solving the second-order differential equation Eq.\ref{eq:diffeq}, the relationship between displacement $y$ and time $t$ can be found. Then, by taking the derivative of $y$, the actual dynamic response acceleration of the free end of the chip under external excitation $a'$ can be obtained. Mechanical stress $\sigma$ can be obtained by dividing the force $Ma'$ generated by the dynamic response of the free end of the crystal by the cross-sectional area of the crystal flake. Subsequently, we can calculate the relationship between stress and time by:
\begin{align}
    \sigma(t)=\frac{Ma'}{bd}. \label{eq:btsolution}
\end{align}

The solution of Eq. \ref{eq:btsolution} should include two parts: one is the forced vibration term generated by external excitation, and the other is the free vibration term of the crystal oscillator in its initial state. For the stress generated by attackers initiating mechanical vibration, we only need to extract the steady-state solution of the forced vibration term:
\begin{align}
    \sigma(t)=\frac{AM^2\omega^2[(M\omega^2-K)\sin(\omega t+\varphi)+S\omega\cos(\omega t+\varphi)]}{bd(K^2-2KM\omega^2+M^2\omega^4+S^2\omega^2)}.
    \label{eq:eq1}
\end{align}

\begin{figure}[h]
\centering
\centerline{\includegraphics[width=0.22\textwidth]{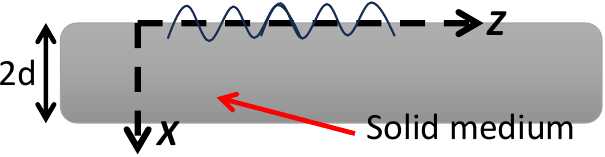}}
\caption{Schematic of sound waves propagating in a $X-Z$ plane with the thickness of $2d$.}
\label{propagation_dic}
\end{figure}

According to Piezoelectric equation\cite{dineva2014piezoelectric}, for the case of vibration only perpendicular to the $X$-direction (see Fig. \ref{propagation_dic}), the electric displacement vector can be denoted as:
\begin{align}
    D=\mu_{33} \cdot \sigma+\epsilon_{33} \cdot E' , \label{eq:eq4}
\end{align}
where $\mu_{33}(>0)$ represents the piezoelectric constant along the $X$-direction, $\epsilon_{33}(>0)$ refers to the dielectric constant along the $X$-direction, and $E'$ represents the external electric field. For the resonance component caused by mechanical vibration, $\epsilon \cdot E'$ can be regarded as 0. Combining Eq. \ref{eq:eq1}$\sim$\ref{eq:eq4}, we can get:
\begin{align}
    D \sim \sigma \sim \sin(\omega t). \label{eq:eqr}
\end{align}
It theoretically demonstrates that the frequency of the electrical signal generated by mechanical vibration is the same as the mechanical vibration frequency $f$. 

\section{Schematic Diagram and Explanation of the Principle of Time Drift Forwards}
\begin{figure}[h]
\centering
\centerline{\includegraphics[width=0.3\textwidth]{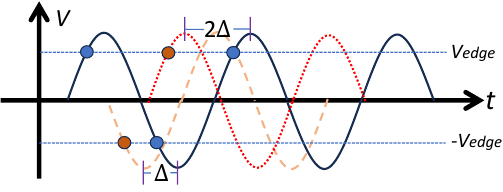}}
\caption{Schematic diagram of the principle of time drift forward.}
\label{samprotation1}
\end{figure}

Under the illustration Fig. \ref{samprotation1}, the black curve represents the initial oscillation signal in the oscillation circuit; The yellow curve represents the phase shift of the oscillation signal forward $\Delta$ after implementing an attack; The red curve indicates that after two attacks, the phase of the oscillation signal drifts forward by $2\Delta$; The blue dots represent the moment when the oscillation signal crosses the rising and falling edge triggering levels three times in a row without implementing an attack; The orange dots represent the moment when the oscillation signal crosses the triggering level of the falling and rising edges during the implementation of two attacks. It can be seen that under the consecutive attack signals, the time when the oscillation signal crosses the edge trigger level is advanced.

\section{Modeling How Timing Drift of the Counter in a BP Monitor Affects the Accuracy of Blood Pressure Measurements}
\label{analysis_bp_monitor}
\begin{figure}[t]
\centering
\includegraphics[width=0.43\textwidth]{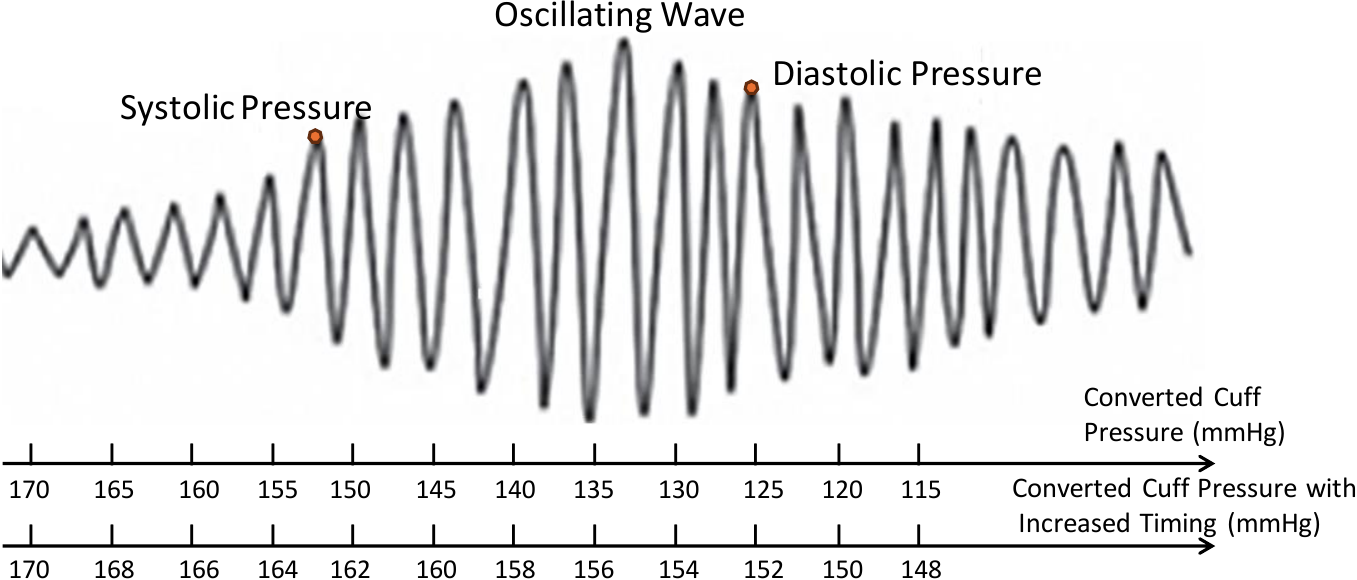}
\caption{Accelerating the rate of cuff deflation causes the oscillation wave to pass through the points corresponding to systolic and diastolic blood pressure in a relatively short period of time, resulting in an overestimation of the cuff pressure value converted at the corresponding time.}
\label{illustration11}
\vspace{-0.1in}
\end{figure}

The BP monitor uses a solenoid valve to apply and release pressure linearly in the cuff. A pressure sensor is applied to detect the amplitude of oscillatory waves during deflation. 
The deflation rate of the solenoid valve, typically 2-3 mmHg/s, is driven by the periodic signal provided by the clock circuit counters. Meanwhile,  
the main controller of the monitor records the amplitude of oscillation waves read by the pressure sensor at different moments. Since the cuff pressure varies uniformly over time, the collected data can be regarded as the amplitudes of oscillation waves at different cuff pressures. Taking the moment of the maximum peak amplitude of the oscillation wave as a reference, the cuff pressure corresponding to 40-45\% of the peak amplitude during the amplitude-increasing phase is identified as the systolic pressure. Conversely, during the amplitude-decreasing phase, the cuff pressure corresponding to 60-75\% of the peak amplitude is identified as the diastolic pressure (see Fig. \ref{illustration11}). The specific ratios are established by the manufacturer's design. Assuming the timing frequency of the counter is $f$ (Hz) and the actual deflation rate of the solenoid valve is $v$ (mmHg/s), with $\Delta P$ being the pressure released by the solenoid valve per clock cycle (mmHg), we have:
\begin{align}
    v=\Delta P \cdot f, \notag
\end{align}
if the timing frequency changes, let the new timing frequency be $f'=f+\Delta f$, the corresponding deflation rate $v'$ will then become:
\begin{align}
    v'=\Delta P(f+\Delta f)=v+\Delta P\cdot \Delta f. \notag
\end{align}
When $\Delta f>0$, the timing rate will become faster than normal, resulting in an increased deflation rate. Conversely, when $\Delta f<0$, the timestamp refresh rate is slower, leading to a decreased deflation rate. 
% In practical measurements, systolic pressure $S$ represents the point at which the oscillatory wave amplitude first exceeds a threshold, corresponding to the time $t_s$ and pressure $P(t_s)$. Diastolic pressure refers to the pressure $P(t_d)$ at which the oscillatory wave amplitude decreases to a specific proportion $k$ of the maximum amplitude, which is determined by the manufacturer's measurement algorithm. 
Given the original deflation rate $V_0$ and the pressure in cuff $P_0$ at the beginning of the measurement, the systolic pressure $S$ can be derived at time $t_s$: $P(t_s)=P_0 -V_0 \cdot t_s=S$. However, given the actual deflation rate $V$, the time at which the cuff pressure reaches the specific point for systolic pressure is $t'_s$:
\begin{align}
    t'_s=\frac{P_0-S}{V_0+\Delta P \Delta f}, \notag
\end{align}
the observed timing drift amount $\Delta t_s$:
\begin{align}
    \Delta t_s=t_s-t'_s=(P_0-S)\cdot \frac{\Delta P \Delta f}{V_0(V_0+\Delta P \Delta f)}. \notag
\end{align}

Therefore, the derived systolic pressure error can be denoted as:
\begin{align}
    \Delta S=V\cdot \Delta t_s=(P_0-S)\cdot \frac{\Delta P \Delta f}{V_0}.
\end{align}

Similarly, the error in diastolic pressure can be derived as $\Delta D=V\cdot \Delta t_d=(P_0-D)\cdot \frac{\Delta P \Delta f}{V_0}$. Thus, when the timing rate increases, the accelerated rate of the cuff deflation causes the peak amplitude of the oscillation wave occurring earlier, leading to an overestimation of the cuff pressure values and, consequently, higher measurements of systolic and diastolic pressures (see Fig. \ref{illustration11}). Conversely, a slower timing frequency delays the peak amplitude occurrence, resulting in underestimations of both systolic and diastolic pressures.

\section{Illustrations and Tables}
\begin{figure}[h]
\centering
\centerline{\includegraphics[width=0.3\textwidth]{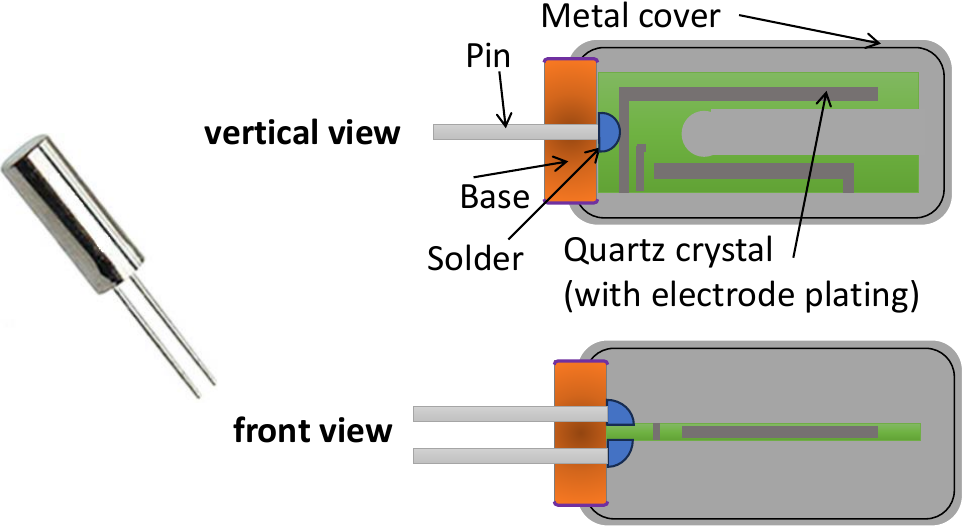}}
\caption{Tuning-fork-shaped quartz crystal structural model.}
\label{crysmanufacture_module}
\end{figure}

\begin{figure}[h]
\centering
\centerline{\includegraphics[width=0.25\textwidth]{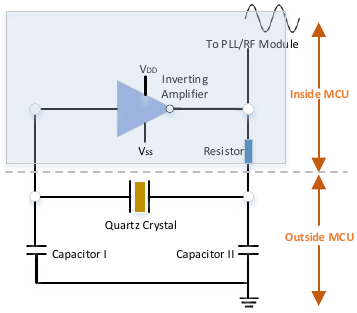}}
\caption{Pierce quartz oscillator module.}
\label{pierceoscillator_module}
\end{figure}

\begin{figure}[!h]
\centering
\centerline{\includegraphics[width=0.22\textwidth]{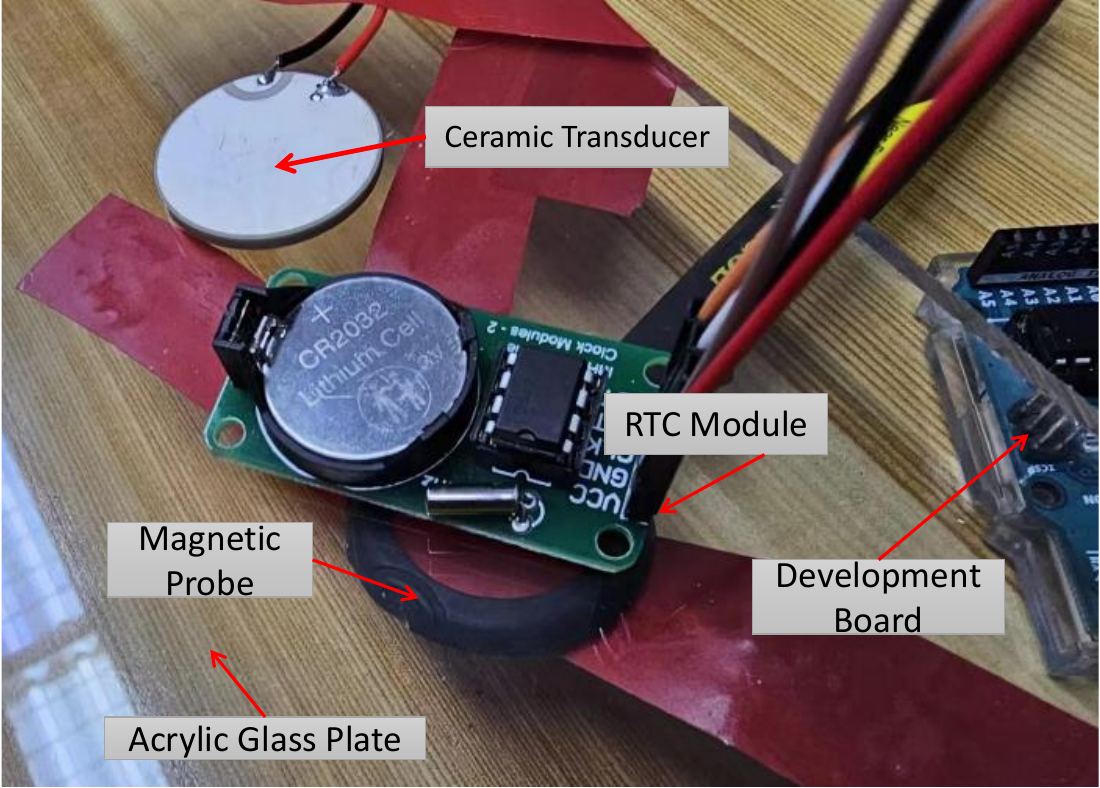}}
\caption{Experimental setup for time drifting backward.}
\label{propagation_solid}
\end{figure}

% \begin{figure}[!h]
% \centering
% \centerline{\includegraphics[width=0.18\textwidth]{pics_and_tables/pic_rotation.pdf}}
% \caption{The placement position of RTC crystal oscillators with different deflection angles relative to the transducer.}
% \label{samprotation}
% \end{figure}

\begin{figure}[!h]
    \centering
    \begin{subfigure}[t]{0.23\textwidth}
        \centering
        \includegraphics[width=\textwidth]{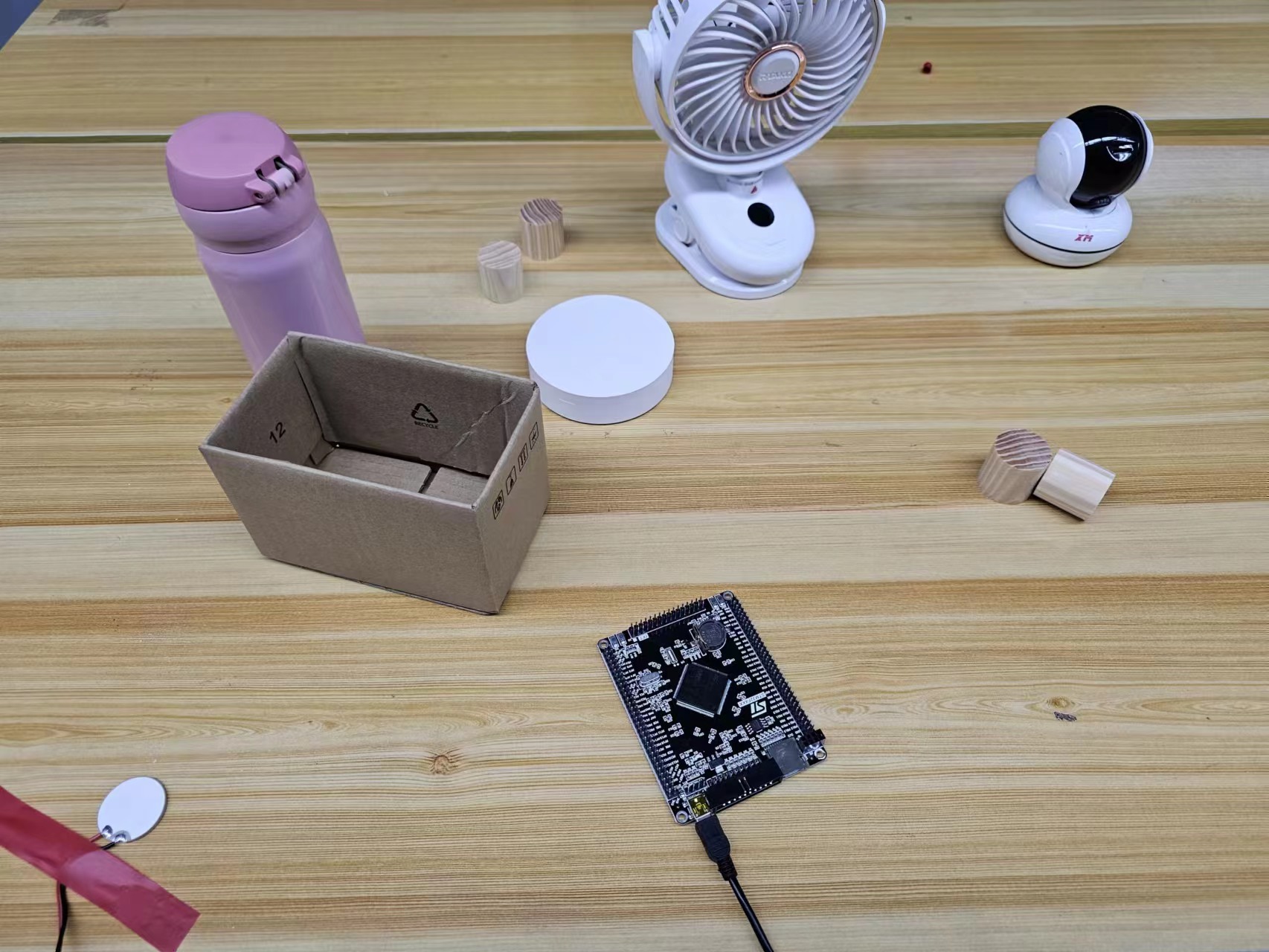}
        \caption{Experimental setting 1.}
    \end{subfigure}
    \hfill % Creates horizontal space between the images
    \begin{subfigure}[t]{0.23\textwidth}
        \centering
        \includegraphics[width=\textwidth]{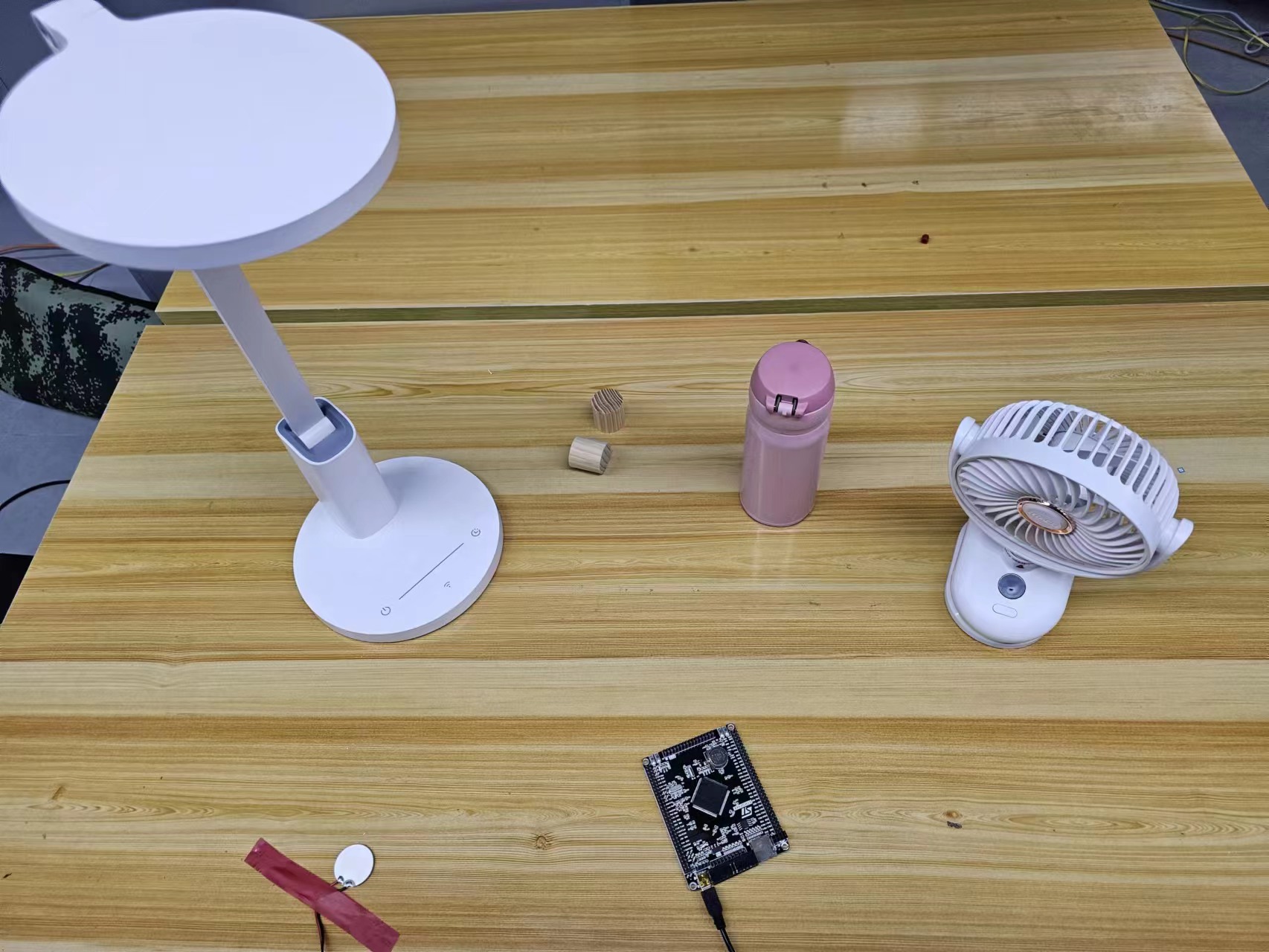}
        \caption{Experimental setting 2.}
    \end{subfigure}
    \hfill
    \caption{Experimental setups on robustness performance test.}
    \label{ac_scenarios_setting}
\end{figure}

% \begin{figure}[!h]
% \centering
% \centerline{\includegraphics[width=0.31\textwidth]{pics_and_tables/pic_21.pdf}}
% \caption{The maximum attack distance under different excitation signal amplitude.}
% \label{tableforwardtest}
% \end{figure}

\begin{figure}[!h]
    \centering
    \begin{subfigure}[t]{0.22\textwidth}
        \centering
        \includegraphics[width=\textwidth]{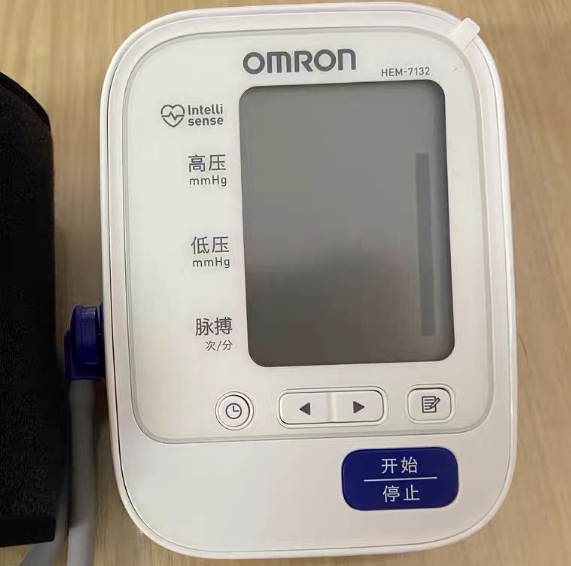}
        \caption{Blood pressure monitor.}
    \end{subfigure}
    \hfill % Creates horizontal space between the images
    \begin{subfigure}[t]{0.15\textwidth}
        \centering
        \includegraphics[width=\textwidth]{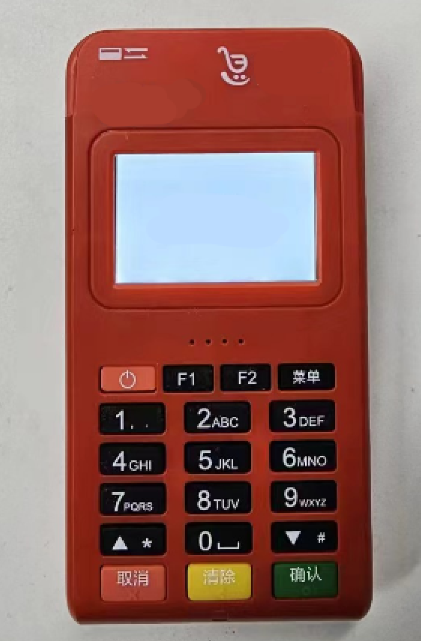}
        \caption{Commercial POS machine.}
    \end{subfigure}
    \hfill
    \caption{Commercial devices under the test.}
    \label{commercial_devices_samples}
\end{figure}

\begin{table}[t]
\centering
\begin{tabular}{|c|c|c|c|}
\hline
\multirow{2}*{\textbf{Medium}} & \multicolumn{2}{c|}{\textbf{Wave Speed($10^3 m/s$)}}& \textbf{Density} \\ \cline{2-3} 
 & \textbf{Longitudinal} & \textbf{Transverse} & ($g/cm^3$)\\ \hline
Aluminum & 6.26 & 3.08 & 2.7\\ \hline
Stainless steel & 6.10 & 3.30 & 7.85 \\ \hline
Quartz glass & 5.57 & 3.52 & 2.2\\ \hline
Acrylic glass & 2.70 & 1.30 & 1.18\\ \hline
Hard rubber & \multirow{2}*{2.30} & \multirow{2}*{0.94} & \multirow{2}*{1.22}\\ 
plastic  &&& \\ \hline
Oak wood & 3.31 & 1.55 & 0.85\\ \hline 
Polyethylene &2.14&0.72& 0.94 \\
\hline
\end{tabular}
\caption{Acoustic transverse and longitudinal wave data in representative solid medium.}
\label{mediumdata}
\end{table}

\begin{table}[t]
\centering
\begin{tabular}{|c|c|c|c|c|}
\hline
\textbf{Module}&\textbf{$\downarrow$5s}&\textbf{$\downarrow$25s}&\textbf{$\uparrow$25s} &\textbf{$\uparrow$45s} \\ \hline
DS1302&$0.81\mu$s&$0.82\mu$s&$0.82\mu$s&$0.8\mu$s\\ \hline
DS1307&$0.79\mu$s&$0.8\mu$s&$0.83\mu$s&$0.81\mu$s\\ \hline
PCF8563T &$0.81\mu$s&$0.82\mu$s&$0.79\mu$s&$0.83\mu$s \\ \hline
DS3231&$0.84\mu$s&$0.78\mu$s&$0.85\mu$s&$0.83\mu$s \\ \hline
STM32F103ZET6 &$0.8\mu$s&$0.79\mu$s&$0.84\mu$s&$0.77\mu$s \\ \hline
XC6C6SLX16&$0.78\mu$s&$0.85\mu$s&$0.8\mu$s&$0.84\mu$s \\ \hline
STM32F103ZGT6 &$0.83\mu$s&$0.81\mu$s&$0.78\mu$s&$0.85\mu$s \\ \hline
STM32F407ZGT6 &$0.86\mu$s&$0.83\mu$s&$0.81\mu$s&$0.78\mu$s \\ \hline
DC-A566 &$0.87\mu$s&$0.85\mu$s&$0.84\mu$s&$0.82\mu$s \\ \hline
\end{tabular}
\caption{Processing delay under different modules with RTC crystal.}
\label{tabledelay}

\end{table}

\begin{table}[!h]
\centering
\begin{tabular}{|c|c|c|c|}
\hline
\textbf{Module}&\textbf{Setting 1}&\textbf{Setting 2}&\textbf{Baseline}  \\ \hline
DS1302&89\%&90\%& 89\%\\ \hline
DS3231&85\%&87\%& 86\%\\ \hline
STM32F10-&\multirow{2}*{82\%}&\multirow{2}*{83\%} & \multirow{2}*{83\%}\\
3ZGT6 &&& \\ \hline
STM32F40-&\multirow{2}*{87\%}
&\multirow{2}*{85\%} & \multirow{2}*{87\%}\\
7ZGT6 &&&\\ \hline
\end{tabular}
\caption{The average attack success rate of representative modules under different environmental background settings.}
\label{res_robustness}
\end{table}

\begin{table}[t]
\centering
\begin{tabular}{|c|c|c|c|}
\hline
\textbf{Probing Distance}&\textbf{Accuracy}&\textbf{Recall}&\textbf{F1 Score}\\ \hline
\multicolumn{4}{|c|}{\textbf{Solid Material: Acrylic Glass}} \\ \hline
\textbf{1cm}&99.43\%& 99.24\% & 99.33\%\\ \hline
\textbf{2cm}&99.41\%& 99.22\% & 99.31\%\\ \hline
\textbf{3cm}&99.38\%& 99.16\% & 99.27\%\\ \hline
\textbf{4cm}&93.29\%& 91.48\% & 92.38\%\\ \hline
\textbf{5cm}&82.43\%& 77.38\% & 79.83\%\\ \hline
\textbf{6cm}&54.69\%& 50.25\% & 52.38\%\\ \hline
\textbf{7cm}&\textcolor{red}{38.74\%}& \textcolor{red}{33.62\%}& \textcolor{red}{36\%}\\ \hline
\multicolumn{4}{|c|}{\textbf{Solid Material: Polyethylene}} \\ \hline
\textbf{1cm}&99.37\%& 99.26\% & 99.31\%\\ \hline
\textbf{2cm}&99.31\%& 99.24\% & 99.27\%\\ \hline
\textbf{3cm}&99.3\%& 99.21\% & 99.25\%\\ \hline
\textbf{4cm}&92.76\%& 90.74\% & 91.74\%\\ \hline
\textbf{5cm}&79.66\%& 72.8\% & 76.08\%\\ \hline
\textbf{6cm}&\textcolor{red}{49.97\%}& \textcolor{red}{47.82\%} & \textcolor{red}{48.87\%}\\ \hline
\textbf{7cm}&\textcolor{red}{31.08\%}& \textcolor{red}{26.65\%}& \textcolor{red}{28.7\%}\\ \hline
\multicolumn{4}{|c|}{\textbf{Solid Material: Hard Rubber Plastic}} \\ \hline
\textbf{1cm}&99.46\%& 99.4\% & 99.43\%\\ \hline
\textbf{2cm}&99.41\%& 99.38\% & 99.39\%\\ \hline
\textbf{3cm}&99.22\%& 98.92\% & 99.07\%\\ \hline
\textbf{4cm}&93.02\%& 91.47\% & 92.24\%\\ \hline
\textbf{5cm}&82.43\%& 80.06\% & 81.23\%\\ \hline
\textbf{6cm}&56.12\%& 52.98\% & 54.5\%\\ \hline
\textbf{7cm}&33.96\%& 29.93\% & 31.82\%\\ \hline
\end{tabular}
\caption{Testing results of classification performance of device models under different detection distances and solid materials.}
\label{modeldetectres}
\end{table}

% \begin{table}[!h]
% \centering
% \begin{tabular}{|c|c|c|c|c|}
% \hline
% \textbf{$\theta_1$}&\textbf{$\theta_2$}&\textbf{DS1302}&\textbf{DS1307} &\textbf{PCF8563T} \\ \hline
% \multirow{3}*{0°}&0°&90\%& 89\%& 87\% \\ 
% &120°& 89\%& 89\% & 88\%\\
% &240°& 90\%& 90\% & 88\%\\ \hline
% \multirow{3}*{60°}&0°&88\% &87\% & 89\%\\ 
% &120°& 90\%& 87\% & 89\%\\
% &240°& 89\%& 88\% & 87\%\\ \hline
% \multirow{3}*{120°}&0°&86\% &89\% & 88\%\\ 
% &120°& 87\%& 89\% & 86\%\\
% &240°& 88\%& 86\% & 90\%\\ \hline
% \multirow{3}*{180°}&0°&86\% &86\% & 88\%\\ 
% &120°& 89\%& 90\% & 88\%\\
% &240°& 88\%& 87\% & 89\%\\ \hline
% \multirow{3}*{240°}&0°&89\% &88\% & 90\%\\ 
% &120°& 89\%& 88\% & 87\%\\
% &240°& 86\%& 87\% & 87\%\\ \hline
% \multirow{3}*{300°}&0°&89\% &87\% & 86\%\\ 
% &120°& 88\%& 89\% & 88\%\\
% &240°& 87\%& 90\% & 89\%\\ \hline
% \end{tabular}
% \caption{The average attack success rate of representative modules at different placement angles from the transducer.}
% \label{impact_directions}
% \end{table}

\begin{table}[!h]
\centering
\begin{tabular}{|c|c|c|}
\hline
\textbf{Layer}&\textbf{Operation}&\textbf{Kernel Size} \\ \hline
1&Input&100000$\times$1 \\ \hline
2&Conv1D&16$\times$64 \\ \hline
3&MaxPool&4 \\ \hline
4&Conv1D&8$\times$128 \\ \hline
5&MaxPool&4 \\ \hline
6&Conv1D&8$\times$256 \\ \hline
7&MaxPool&4 \\ \hline
8&Flatten&- \\ \hline
9&FC1&1024 \\ \hline
10&Dropout&0.2 \\ \hline
11&Output+Softmax &16* \\ \hline
\end{tabular}
\caption{Neural network structure for model classification. The output classification labels include 15 device models as well as an ``unidentified" label.}
\label{classifynetwk}
\end{table}

\end{document}